\documentclass[aps,twocolumn,prb,superscriptaddress,footinbib,floatfix]{revtex4-2}
\usepackage{amsmath,amssymb,bm}
\usepackage{graphicx}
\usepackage{epstopdf}
\usepackage{latexsym}
\usepackage[normalem]{ulem} 
\usepackage{appendix}
\usepackage{dsfont}
\usepackage[usenames,dvipsnames]{color}
\usepackage{hyperref}
\usepackage{natbib}
\usepackage[dvipsnames]{xcolor}
\usepackage{braket}
\usepackage{mathtools}
\usepackage{empheq}
\usepackage{mdframed}
\usepackage{tcolorbox}
\usepackage{environ}
\NewEnviron{equationS}{
\begin{align}\begin{split}
  \BODY
\end{split}\end{align}
}
\begin{document}

\newcommand{\Rev}[1]{{\color{blue}{#1}\normalcolor}} 
\newcommand{\Com}[1]{{\color{red}{#1}\normalcolor}} 
\newcommand*\widefbox[1]{\fbox{\hspace{2em}#1\hspace{2em}}}

\newcommand{\ketbra}[2]{\ket{#1}\hspace{-0.12cm}\bra{#2}}
\newcommand{\normord}[1]{\mathopen{:}\,#1\,\mathopen{:}}
\newcommand{\nocontentsline}[3]{}
\newcommand{\tocless}[2]{\bgroup\let\addcontentsline=\nocontentsline#1{#2}\egroup}

\author{D. Barberena}
\affiliation{JILA, Department of Physics, University of Colorado,  Boulder, CO 80309, USA}
\affiliation{Center for Theory of Quantum Matter, University of Colorado, Boulder, CO 80309, USA}
\affiliation{T.C.M. Group, Cavendish Laboratory, University of Cambridge, J.J. Thomson Avenue, Cambridge CB3 0HE, UK}
\author{Matthew P. A. Fisher}
\affiliation{Department of Physics, University of California, Santa Barbara, CA 93106}

\title{Postselection in lattice bosons undergoing continuous measurements}

\date{\today}

\begin{abstract}
We study in detail the postselection problem in a specific model: bosons hopping on a lattice subjected to continuous local measurements of quadrature observables. We solve the model analytically and show that the postselection overhead can be reduced by postprocessing the entire measurement record into one or two numbers for each trajectory and then postselecting based only on these numbers. We then provide a step-by-step protocol designed to recover connected two-point functions of the quantum trajectories, which display an exponentially decaying profile that is not observable in the unconditional, trajectory averaged, state. With the analytical solution in hand, we analyse the features of this postprocessing stage with the intention of abstracting away the properties that make postselection feasible in this model and may help in mitigating postselection in more general settings. We also test the protocol numerically in a way that utilizes only experimentally accessible information, showing that various quantum trajectory observables can be recovered with a few repetitions of the numerical experiment, even after including inevitable coarse-graining procedures expected under realistic experimental conditions. Furthermore, all the information required to design the postprocessing stage is independently present both in the unconditional dynamics and also in the measurement record, thus bypassing the need to solve for the conditional evolution of the model. We finalize by providing experimental implementations of these models in cavity-QED and circuit-QED.
\end{abstract}

\maketitle  

\section{Introduction}

Measurements in quantum mechanics are like a bargain with the devil: in exchange for a reduced amount of information, we perturb the measured physical system in a fundamentally unavoidable, often destructive, way. Nevertheless, carefully designed measurement strategies can exploit the little information we gain to then prepare entangled/non-classical quantum states by means of e.g. quantum non-demolition approaches~\cite{Braginsky1996,Kuzmich2000,SchleierSmith2010,Cox2016,Hosten2016} or heralding~\cite{Knill2001,Nielsen2004,Moehring2007,Cao2024}, making measurements an important resource for modern quantum technologies. This is a capability that has been long recognized in the study of monitored quantum systems~\cite{belavkin2005,Bergquist1986,Korotkov1999,Jacobs2006,Carmichael2008,Wiseman_Milburn_2009,Clerk2010,Minev2019}, and over the years many works have focused on tractable few-body models for the purposes of estimation and sensing tasks~\cite{Caves1980,Mabuchi_1996,Tsang2012,Gammelmark2014}.

Quantum mechanical measurements can also give rise to unexpected phenomena in many-body dynamics~\cite{Li2018,Chan2019,Skinner2019,annurevFisher}. In this different context, the competition between local projective measurements and random unitary gates was found to lead to a transition in the entanglement structure of the system under study, from an area law phase where measurements dominate, to a volume law phase where unitary evolution efficiently scrambles information. Since this discovery, many efforts have been devoted to elucidate the role played by measurements in the many-body setting~\cite{Gullans2020,Bao2020,Ippoliti2021,Alberton2021,Minoguchi2022,Poboiko2024} and how they fit into our understanding of quantum phases of matter~\cite{Friedman2023}. In particular, there has been a lot of emphasis on characterization by means of quantum information probes such as von-Neumann and Renyi entropies, and mutual informations, which provide basis-independent diagnostics of entanglement and correlations.

At the same time, many-body systems undergoing measurement dynamics suffer from the so-called ``postselection problem". This problem is the consequence of two facts: (1) the quantum state $\ket{\psi}$ of a system that has been subject to measurement+unitary dynamics depends (in principle) on the outcomes of all the measurements that were performed during its evolution history; (2) to be able to extract information from $\ket{\psi}$ by means of statistics (to obtain, e.g., the variance of an observable) one needs to build an ensemble of identical copies of $\ket{\psi}$. Characterizing $\ket{\psi}$ thus requires repeating the same measurement+unitary dynamics with the same outcomes for all the measurements. Since the result of a quantum mechanical measurement is non-deterministic, in most cases the outcomes will not be the same. In fact, for extended systems and evolutions with an extensive number of measurements, the probability of obtaining the same set of outcomes is exponentially small in system size and evolution time. Thus, building an ensemble of copies of $\ket{\psi}$ (i.e. the ``postselected ensemble") becomes impossible as a matter of principle, except for very small systems. Because of this, experimental observations of measurement-induced transitions are challenging~\cite{Noel2022,Koh2023,Hoke2023}, even after overlooking the inherent complications of measuring entropies~\cite{Sackett2000,Kaufman2016,Kokail2021,Joshi2023} for large systems.

To address this singular complication, which can put into question the identification of measurement-induced dynamical phases as genuine phases of matter~\cite{Friedman2023}, further research has focused on discerning circumstances in which the complexities introduced by postselection can be partially mitigated: circuits with space-time duality~\cite{Ippoliti2021b,Lu2021}, the use of quantum-classical hybrid approaches in cases where classical simulation is feasible~\cite{Li2023,McGinley2024,Garratt2024}, fully connected models where only a small subregion of Hilbert space is explored~\cite{Passarelli2024}, systems in the semiclassical regime~\cite{li2024,delmonte2024}, etc. In fact, the ability to overcome postselection in an efficient way may be an important characterization on its own, possibly related to the ability to use the measurement record to learn information of the quantum state that is subjected to the hybrid unitary+measurement dynamics~\cite{Dehghani2023,Akhtar2024,Ippoliti2024}. Moreover, applying feedback operations conditioned on the measurement outcomes has been shown to lead to effects that can be observed without postselection~\cite{Ivanov2020,Ivanov2021,Friedman2023,Piroli2023,Ravindranath2023,wang2023,Odea2024,Hauser2024}, although the connection to the original evolution without feedback is tenuous because the extra applied operations strongly modify the dynamics of the system in question~\cite{Ravindranath2023,Piroli2023,Odea2024}. 

In this paper we study the structure of postselection in an analytically solvable model that is also experimentally realizable using modern quantum platforms: bosons hopping on a lattice subjected to local continuous measurements of quadrature observables. Instead of focusing on entropies, we investigate local observables, for which the requirements on postselection are loosened drastically. This is because local observables in a given quantum trajectory are reconstructed by:
\begin{enumerate}
    \item Distilling the macroscopic number of measurement results into a few numbers (one or two), called ``estimators"~\cite{Jacobs_2014}.
    \item Binning quantum trajectories together according to whether the estimators acquire the same value, as opposed to the entire measurement record.
\end{enumerate}
Importantly, the estimators depend on the choice of observable. In practice, the post-selection problem is still present because determining the relation between the estimators and the measurement record is generally a hard task, but in our model we can construct them explicitly. We show that:
\begin{itemize}
    \item The estimators display structure consistent with spatial locality, in the following sense: if we want to reproduce two-point correlation functions nearby locations $\mathbf{r,r'}$ and at an observation time $T$, the relevant estimators are constructed from the measurement record in the vicinities of $\mathbf{r,r'}$ and close to the time $T$. This is a consequence of the emergence of a correlation length $\xi$ and an associated memory time $\tau$ as a result of the competition between measurements and unitary evolution~\cite{Gullans2020b,Minoguchi2022}. Unlike the case of random circuits, $\xi$ and $\tau$ are not statistical~\cite{Gullans2020b,Ippoliti2021}, but are properties of each trajectory. In our lattice model we can tune the parameters to make the correlation length much larger than the lattice spacing, leading to an effective continuous description.
    \item The procedure to construct the estimators in terms of the measurement record can be inferred directly from the measurement record itself, or by an adequate analysis of the unconditional dynamics, i.e. after averaging over all possible measurement realizations~\cite{Muller2009}. Although this is true in general, the problem of reconstructing the estimator is substantially simplified if there is some prior knowledge about the functional relation between the measurement record and the estimator.
\end{itemize} 
We also verify these results by performing numerical experiments that mimic actual experimental conditions, where after a single iteration of an experiment the only outcomes are the measurement record and a single copy of the quantum state. This procedure further illustrates that the conditional evolution needs to be repeated only a few times ($\sim 10^4$ is sufficient for our model), and that many different observables can be recovered using the same set of quantum trajectories. 

Quadratic bosonic models in the context of measurement transitions have been investigated in the past~\cite{Minoguchi2022,wang2023,yokomizo2024}, and fall within the scope of Kalman-Bucy filter theory~\cite{Kalman1960,BELAVKIN1999,Jacobs_2014}. Although they are analytically solvable, some of their phenomenology, such as the emergence of a memory time, extends beyond the setting of quadratic bosons~\cite{lin2023}. This may very well provide some mitigation of postselection in some situations. We thus approach our model with the explicit intention of identifying features that may generalize beyond the quadratic boson realm.

This paper is organized as follows: 
\begin{description}
    \item[Section~\ref{sec:Def}] We introduce the lattice model and commment on its basic features.
    \item[Section~\ref{sec:SingleSite}] We study the system in a single lattice site. The solution in this simplified setting will guide the investigation of the full lattice system. Here we describe a step-by-step procedure to recover variances of observables. We also show that this procedure can be designed based only on information present in the measurement record and/or the unconditional dynamics. 
    \item[Section~\ref{sec:Multisite}] We embark on the analysis of the lattice system introduced in Section~\ref{sec:Def} and provide fully analytical solutions for its stationary steady state. We show that the postselection procedure introduced in Section~\ref{sec:SingleSite} works in this setting too. We also discuss some features of postselction as the continuum limit is approached.
    \item[Section~\ref{sec:ExpImp}] We describe possible experimental implementations of the models studied in the previous sections.
\end{description}

\section{Definition of the model}\label{sec:Def}
\begin{figure}
    \centering
    \includegraphics[width=0.4\textwidth]{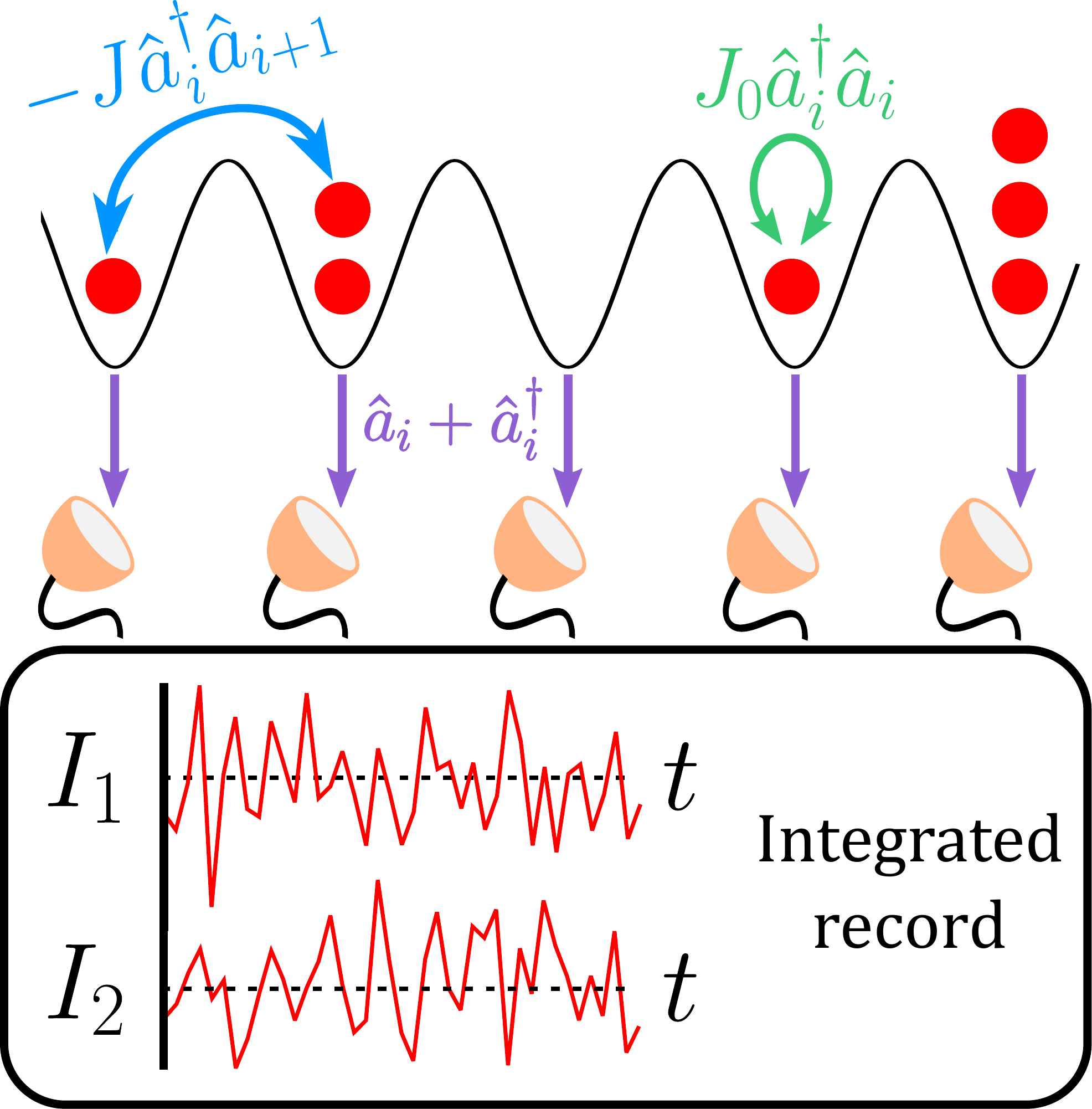}
    \caption{Schematic of the lattice system. Bosons (red circles) hop around with hopping matrix element $-J$ and onsite energy $J_0$. Each lattice site is subjected to measurements of $\hat{a}_i+\hat{a}_{i}^\dagger$. The output of the measurements is a sequence of numbers that can be integrated to give the integrated record $I_i$ at each lattice site, as depicted in the figure.}
    \label{fig:Schematic}
\end{figure}

The system is also subjected to on-site measurements of the quadrature operators $\hat{x}_i=(\hat{a}_i+\hat{a}^\dagger_i)/\sqrt{2}$. In the photon context, these correspond to measurements of the electric or magnetic field. We model the measurements using the stochastic Schr\"odinger equation formalism~\cite{Jacobs_2014,Wiseman_Milburn_2009,ALBARELLI2024}, which describes the conditional evolution of a state $\hat{\rho}$ undergoing joint coherent and measurement dynamics. This is a stochastic differential equation, whose Ito form is given by
\begin{equationS}\label{eqn:Def:SSE}
    d\hat{\rho}&=-i\left[\hat{H},\hat{\rho}\right]\,dt+\Gamma\sum_i\left(\hat{x}_i\hat{\rho}\,\hat{x}_i-\frac{1}{2}\{\hat{x}_i^2,\hat{\rho}\}\right)\,dt\\
    &+\underbrace{\sqrt{\Gamma}\sum_i\Big(\hat{x}_i\hat{\rho}+\hat{\rho}\hat{x}_i-2\braket{\hat{x}_i}\hat{\rho}\Big)\,dW_i}_{\text{stochastic}}.
\end{equationS}
To preserve translational invariance, we assume that all the local measurements have the same strength $\Gamma$. The first line in Eq.~(\ref{eqn:Def:SSE}) is deterministic, and consists of a coherent term $\propto \hat{H}$ and measurement induced terms $\propto \Gamma$ that have the form of a Lindblad superoperator. The second line in Eq.~(\ref{eqn:Def:SSE}) is stochastic, as evidenced by the presence of the Wiener increments $dW_i$, and models the non-deterministic nature of quantum mechanical measurement outcomes and their back-action on the measured quantum state. The presence of the deterministic Lindblad term reflects the intimate relationship that exists between measurements and dissipation. If we were to average over different realizations of the measurement process, the second line disappears ($\overline{dW_i}=0$, where the bar indicates averaging over measurement realizations), and we are left with standard open system dynamics for the unconditional quantum state $\underline{\hat{\rho}}$. In what follows, we will always refer to measurement-averaged quantities (i.e. unconditional) using an overline or an underline.

The dynamical evolution of the quantum state given by Eq.~(\ref{eqn:Def:SSE}) must be supplemented by an equation for the measurement record
\begin{equation}
    dI_i=2\sqrt{\Gamma}\braket{\hat{x}_i}\,dt+dW_i,
\end{equation}
which consists of a sequence of increments $\big[dI_i(0),\,...\,,\,dI_i(t)\big]\equiv\{dI_i\}$ for each time step at each lattice site $i$. Importantly, experimentalists only have access to $dI_i$, not to $dW_i$, so the actual outputs of an experiment are the record $\{dI_i\}$ and the conditional quantum state $\hat{\rho}$, which \textit{a priori} depends on all the $dI_i$. Nevertheless, the Wiener increments $dW_i$ are useful for numerical solution of the coherent+measurement dynamics since they are known to follow a zero mean Gaussian distribution with a variance of size $dt$. It is worth pointing out that the record $\{dI_i\}$ is something that is often integrated, so expressions like $\int f(t)dI_i(t)$ (for some function $f$) will be common. For instance, Fig.~\ref{fig:Schematic} shows the integrated record $I_i(t)=\int_0^t dI_i(s)$, which is continuous and hence easier to depict graphically than $dI_i(s)$, which is everywhere discontinuous.

In a given quantum trajectory, the system will develop correlations that can be characterized by connected two-point functions like
\begin{equationS}\label{eqn:Def:CP}
    C^X_{ij}&=\braket{\hat{x}_i\hat{x}_j}-\braket{\hat{x}_i}\braket{\hat{x}_j}\\
    C^P_{ij}&=\braket{\hat{p}_i\hat{p}_j}-\braket{\hat{p}_i}\braket{\hat{p}_j},
\end{equationS}
where $\hat{p}_i=(\hat{a}_i-\hat{a}_i^\dagger)/\sqrt{2}$. Crucially, these correlators are nonlinear in the conditional quantum state $\hat{\rho}$ because of the subtraction of the unconnected components, i.e. $\braket{\hat{p}_i}\braket{\hat{p}_j}$, and are thus not accessible to the unconditional quantum state $\underline{\hat{\rho}}$. They are ``hidden" behind the postselection barrier. Measuring them would naively require us to repeat the experiment many times and hope that the increments $dI_i$ at every single time step and lattice point return the same value. This is, of course, out of the question. Our task in this paper will be to find efficient ways of doing postselection to access observables like $C^{X,P}_{ij}$ without requiring that all the increments $dI_i$ be the same. We will do this by carefully analyzing the analytical solution of the system dynamics, by which we mean an explicit expression for $\hat{\rho}$ as a function of time and of the increments $dI$. We will also complement our analytical arguments with numerics that closely parallel real experiments. 

We finish this section by pointing out a number of important features of the model defined by Eq.~(\ref{eqn:Def:SSE}): (i) boson number $\sum_i\hat{a}_i^\dagger\hat{a}_i$ is not conserved because of the measurements, which can arbitrarily create or destroy particles, (ii) the generator of time translations is quadratic in boson operators, so that gaussian initial states remain gaussian for all time, (iii) even generic non-gaussian states become gaussian at longer times, so that a gaussian description is always appropriate (this is demonstrated in Section~\ref{subsec:SSite:Nongaussian}).   

\section{Single site model}\label{sec:SingleSite}
In this section we consider a single site version of Eq.~(\ref{eqn:Def:SSE}) 
\begin{equationS}\label{eqn:SSite:SSE}
    d\hat{\rho}&=-i\left[h_0\hat{a}^\dagger\hat{a},\hat{\rho}\right]\,dt+\Gamma\left(\hat{x}\hat{\rho}\,\hat{x}-\frac{1}{2}\{\hat{x}^2,\hat{\rho}\}\right)\,dt\\
    &+\sqrt{\Gamma}\Big(\hat{x}\hat{\rho}+\hat{\rho}\hat{x}-2\braket{\hat{x}}\hat{\rho}\Big)\,dW
\end{equationS}
where the Hamiltonian only has the on-site energy term $h_0\hat{a}^\dagger\hat{a}$. This model, variants and generalizations have been studied extensively in the literature~\cite{Doherty1999,Muller2009,Clerk2010,Jacobs_2014,Meng2020}, with a lot of emphasis on estimation, amplification and feedback. It is particularly relevant in the field of cavity optomechanics~\cite{Clerk2010,Aspelmeyer2014}, in which $\hat{x}$ is a proper position operator. Here we will focus instead on those features that will be relevant for the lattice model and postselection. Before proceeding, let us also reiterate some of the points discussed in the previous section:  the measurement record consists of a sequence of increments $dI$ at each time $t$, calculated according to
\begin{equation}\label{eqn:SSite:Record}
    dI=2\sqrt{\Gamma}\braket{\hat{x}}\,dt+dW,
\end{equation}
the output of the measurement process consists of the record $\{dI\}$ and a single copy of the conditional quantum state $\hat{\rho}$, and the evolution preserves gaussianity. 

\begin{figure}
    \centering
    \includegraphics[width=0.48\textwidth]{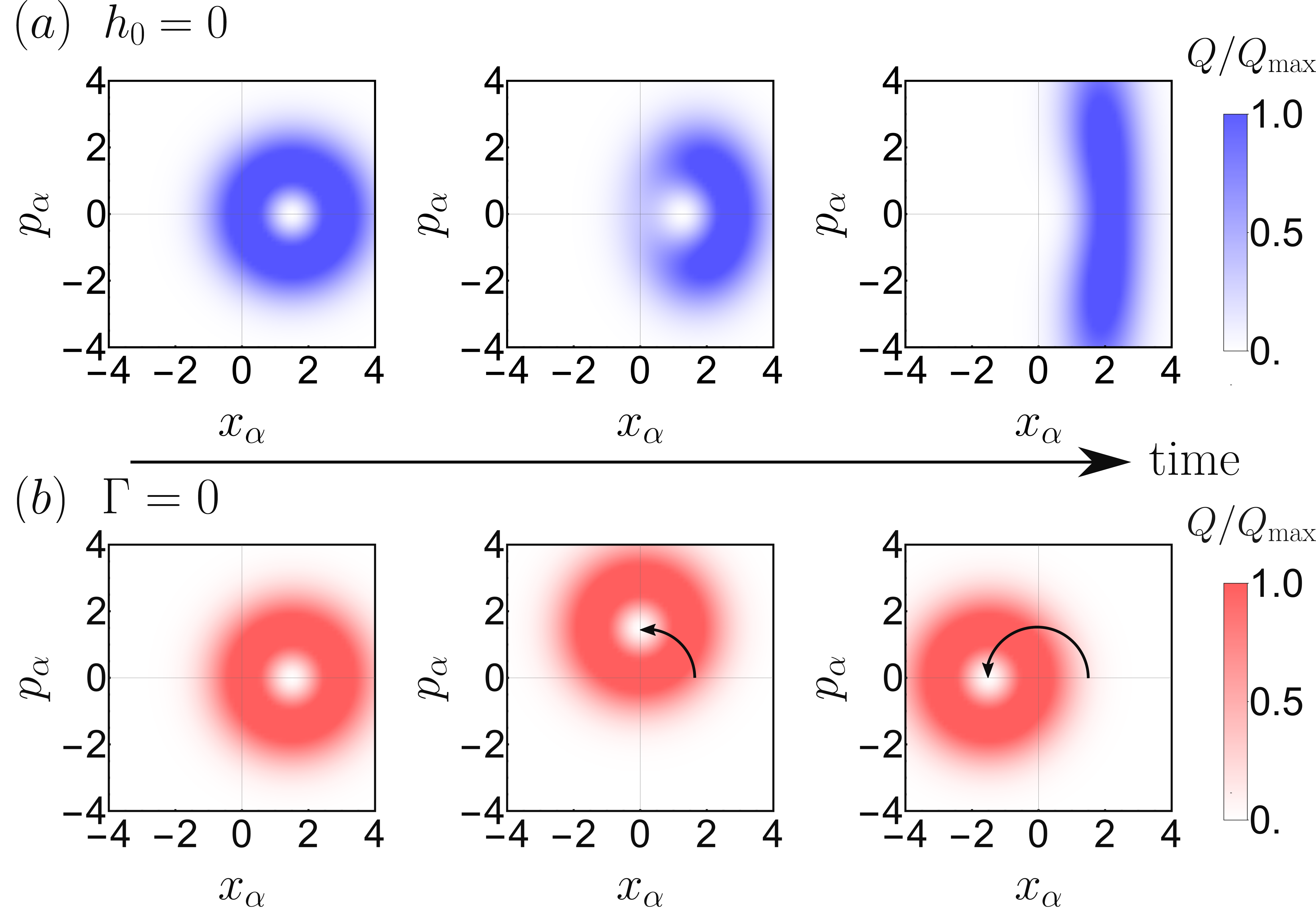}
    \caption{Husimi probability distributions $Q(\alpha)=\braket{\alpha|\hat{\rho}|\alpha}/\pi$ for (a) only measurements and (b) only unitary evolution as a function of time.}
    \label{fig:SingleSite:IndependentEvolutions}
\end{figure}

Measurements alone ($h_0=0$) would drive the system towards an eigenstate of $\hat{x}$, although this would only be attained at infinite times. At finite times, we would observe that fluctuations in $\hat{x}$ (as measured by e.g. the variance) steadily decrease, reaching $0$ as $t\to \infty$. Because of the uncertainty principle, fluctuations in $\hat{p}$, the conjugate variable, must necessarily increase without bound, reaching $\infty$ as $t\to \infty$. We depict this process graphically using Husimi quasiprobability distributions in phase space, defined by $Q(\alpha)=\braket{\alpha|\hat{\rho}|\alpha}/\pi$ ($\ket{\alpha}$ is a coherent state) and plotted as a function of $x_\alpha=\sqrt{2}\Re(\alpha)$ and $p_{\alpha}=\sqrt{2}\Im (\alpha)$. Fig.~\ref{fig:SingleSite:IndependentEvolutions}(a) shows the fate of a displaced one boson state, whose extent along the $x_{\alpha}$/$p_\alpha$ direction becomes progressively smaller/larger.

The onsite energy alone ($\Gamma=0$) describes familiar harmonic oscillator dynamics. Its action in phase space is easy to interpret: it rotates the distribution $Q(\alpha)$ without distorting it [see Fig.~\ref{fig:SingleSite:IndependentEvolutions}(b)]. When both terms are present, measurements try to squeeze the distribution along the $x_\alpha$ direction, but the onsite energy will rotate the squeezed direction, preventing relaxation towards a $\hat{x}$ eigenstate. If $h_0\gg \Gamma$, the rotation is so fast that any attempt to squeeze the distribution should be quickly washed out. If $h_0\ll \Gamma$ a fair amount of distortion should be possible, but the rotation will eventually cap it out. On top of this, measurement backaction leads to a stochastic drift of the location of the distribution~\cite{Doherty1999,Meng2020}.

To make the following discussion simpler, we will first focus on gaussian initial states and discuss the more general case at the end of the section. Because of the gaussianity assumption, we can characterize the quantum state entirely in terms of its means $\braket{\hat{x}}$, $\braket{\hat{p}}$ and covariances 
\begin{equationS}
    v_x&=\braket{\hat{x}^2}-\braket{\hat{x}}^2\\
    v_p&=\braket{\hat{p}^2}-\braket{\hat{p}}^2\\
    u&=\frac{\braket{\hat{x},\hat{p}}}{2}-\braket{\hat{x}}\braket{\hat{p}}.
\end{equationS}
In this single site setting, the covariances play the role of $C^X_{ij},C^P_{ij}$ from Eq.~(\ref{eqn:Def:CP}) because they are nonlinear in the conditional quantum state $\hat{\rho}$.
\subsection{Mathematical solution}
Simple evolution equations for the means and covariances can be obtained directly from Eq.~(\ref{eqn:SSite:SSE}) after using gaussianity to get rid of higher order correlators. As is typical for these systems, the equations for the covariances are nonlinear, deterministic, and close among themselves (see Appendix~\ref{app:SSite:Gaussian} or Ref.~\cite{Jacobs_2014}):
\begin{equationS}\label{eqn:SSite:Covs}
    \dot{v}_x&=2h_0 u-4\Gamma v_x^2\\
    \dot{v}_p&=-2h_0 u+\Gamma-4\Gamma u^2\\
     \dot{u}&=h_0(v_p-v_x)-4\Gamma u v_x,
\end{equationS}
while the equations for the means depend on the covariances and are stochastic (depend on $dW$ or $dI$)
\begin{equation}\label{eqn:SSite:Means}
      d\begin{pmatrix}
        \braket{\hat{x}}\\ \braket{\hat{p}}
    \end{pmatrix}=\begin{pmatrix}
        -4\Gamma v_x&h_0\\-h_0-4\Gamma u&0
    \end{pmatrix}\begin{pmatrix}
        \braket{\hat{x}}\\ \braket{\hat{p}}
    \end{pmatrix}\,dt+2\sqrt{\Gamma}\begin{pmatrix}
        v_x\\
        u
    \end{pmatrix}\,dI,
\end{equation}
where we have used Eq.~(\ref{eqn:SSite:Record}) to eliminate $dW$ in favor of $dI$ because, as mentioned before, experiments only have access to $dI$.
\begin{figure}
    \centering
    \includegraphics[width=0.48\textwidth]{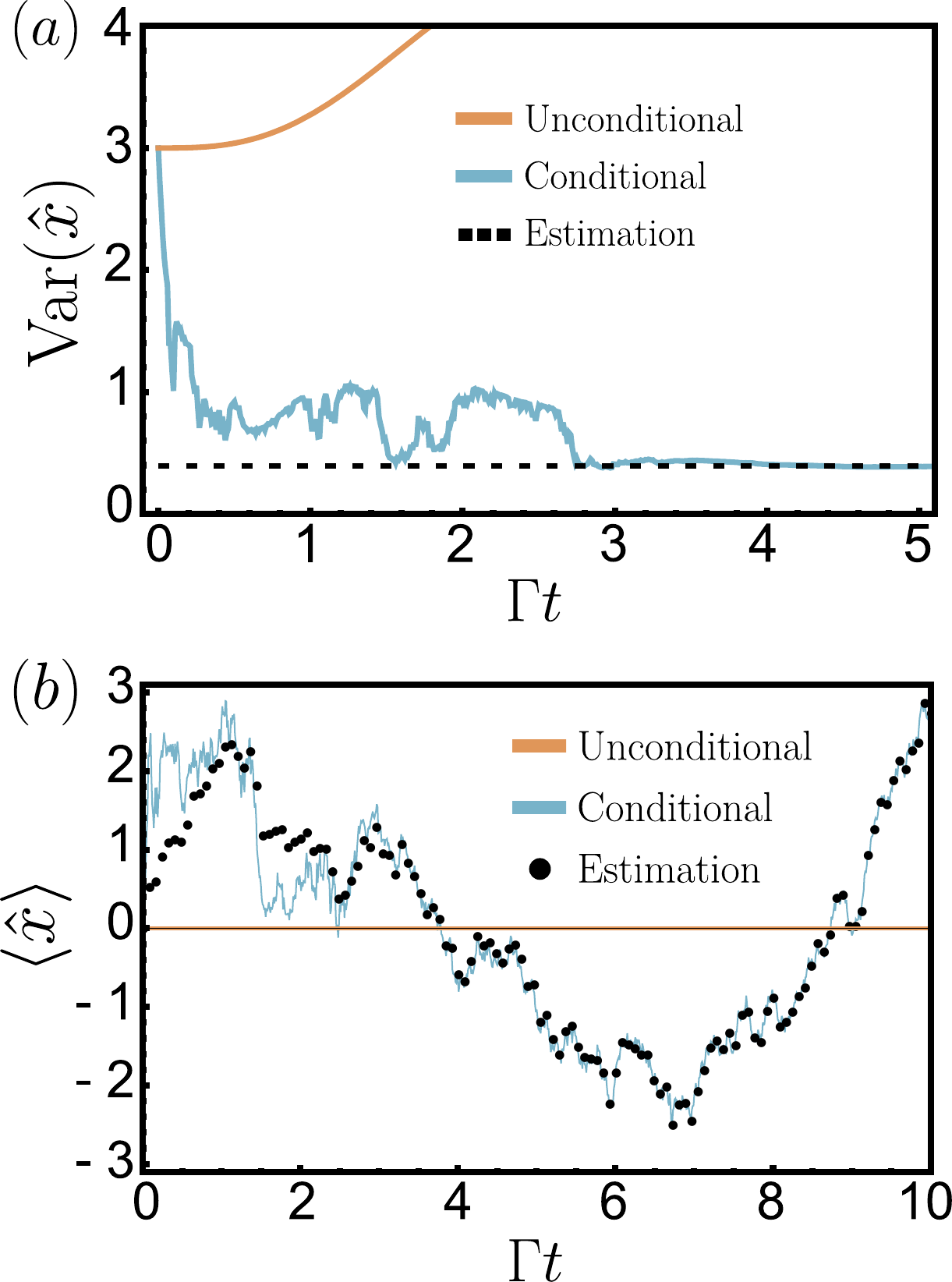}
    \caption{(a) Variance and (b) mean of $\hat{x}$ in a single quantum trajectory as a function of time from $t=0$ to $t=10\Gamma^{-1}$ for $h_0=\Gamma$. The initial state was $(\ket{0}+\ket{5})/\sqrt{2}$. Blue lines are directly calculated from the time-evolved quantum state [using Eq.~(\ref{eqn:SSite:SSE})], black lines or dots are the analytical formulas given in Eq.~(\ref{eqn:SSite:VarianceSolution}) and Eq.~(\ref{eqn:SSite:MeanEstimator}), and orange lines represent the unconditional results of Eq.~(\ref{eqn:SSite:UnconditionalObservabes}).}
    \label{fig:SingleSite}
\end{figure}

The system of equations given in Eq.~(\ref{eqn:SSite:Covs}) has a unique stable fixed point so that, at long times, we have 
\begin{equation}\label{eqn:SSite:VarianceSolution}
    v_x^{\infty}=\frac{\sqrt{\frac{h_0}{8\Gamma}}}{\left(\frac{h_0}{4\Gamma}+\sqrt{\left(\frac{h_0}{4\Gamma}\right)^2+\frac{1}{4}}\right)^{1/2}}
\end{equation}
independently of the initial conditions of the system. When measurements are strong, i.e. $\Gamma\gg h_0$, then $v_x^{\infty}\to 0$ because the system is driven towards a nearly perfect $\hat{x}$ eigenstate. In the opposite regime, $h_0\gg \Gamma$, coherent dynamics continuously mixes the $\hat{x}$ and $\hat{p}$ variances and averages out the effects of measurements, as discussed before.

At these long times, the system of equations for the means [Eq.~(\ref{eqn:SSite:Means})] is time-translationally invariant and can be solved in terms of exponentials. In particular, the late time value of $\hat{x}$ can be estimated via $x_{\text{est}}$, defined by
\begin{equation}\label{eqn:SSite:MeanEstimator}
    \frac{x_{\text{est}}}{2\sqrt{\Gamma} v_x^{\infty}}=\int_{-\infty}^Te^{-2\Gamma v_x^{\infty}(T-s)}\cos\left[\frac{h_0(T-s)}{2v_x^{\infty}}\right] dI(s),
\end{equation}
where we are being explicit about the time label on $dI(s)$. The measurement record $dI(t)$ must be ``filtered" via a filter function that in this case turns out to be oscillatory with exponential decay. In Fig.~\ref{fig:SingleSite} we show a comparison between numerical simulation of Eq.~(\ref{eqn:SSite:SSE}) and the analytical estimates given by Eq.~(\ref{eqn:SSite:VarianceSolution}) and Eq.~(\ref{eqn:SSite:MeanEstimator}). After an initial transient, we verify that the analytical formulas correctly capture the long time behaviour of means and variances. Note that the initial state in Fig.~\ref{fig:SingleSite} was chosen to be $\propto \ket{0}+\ket{5}$, where $\ket{n}$ is the state with $n$ bosons. This is a markedly non-gaussian state but the late time dynamics is insensitive to this, as we will show at the end of the section. 

\begin{figure}
    \centering
    \includegraphics[width=0.48\textwidth]{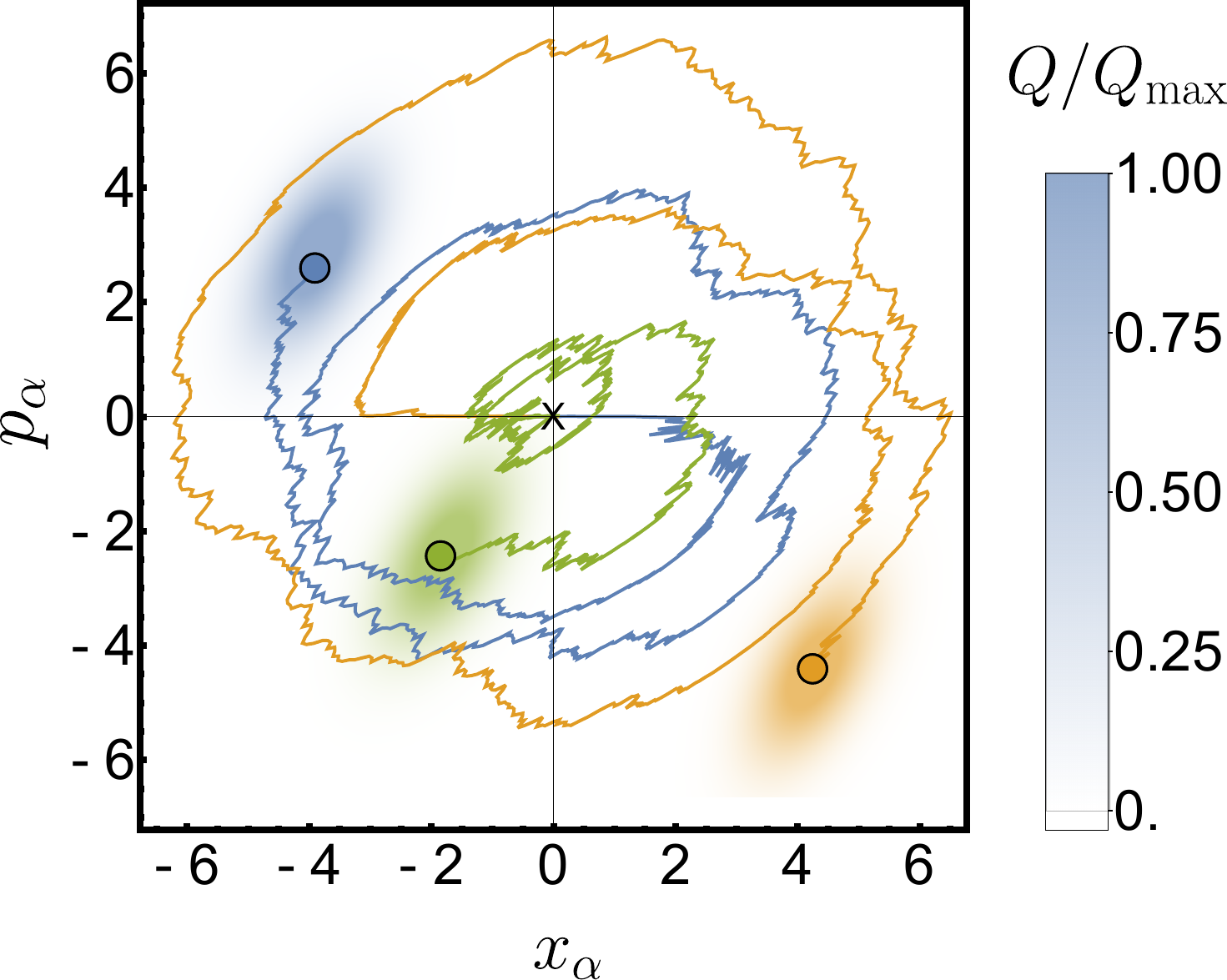}
    \caption{Husimi quasiprobability distributions [$Q(\alpha)=\braket{\alpha|\hat{\rho}|\alpha}/\pi$] for three different realizations of the measurement dynamics beginning from $\ket{0}$. We parameterize $\alpha$ in terms of its real and imaginary parts $\alpha=(x_\alpha+ip_\alpha)/\sqrt{2}$ and normalize $Q$ to its maximum value $Q_{\text{max}}$. The three distributions are located at different points in phase space, but their shape and orientation is identical. The solid lines indicate the time evolution of the centroid of each blob.}
    \label{fig:SingleSite:Husimi}
\end{figure}

These results can be depicted graphically by using Husimi distributions. Fig.~\ref{fig:SingleSite:Husimi} shows three different realizations of the measurement dynamics that start from the same vaccuum $\ket{0}$ state: as time progresses, the centroid of the distribution evolves stochastically, but its shape and orientation (which depend on $v_x$, $v_p$ and $u$) become fixed in time. In different quantum trajectories, the centroid will end up in various different locations, so the unconditional quantum state will have a continuously growing distribution with $\overline{\braket{\hat{x}}}=0$ and variance given by (see Fig.~\ref{fig:SingleSite:Husimi})
\begin{equationS}\label{eqn:SSite:UnconditionalObservabes}
    \overline{\braket{\hat{x}^2}}-\left(\overline{\braket{\hat{x}}}\right)^2&=\frac{1+\Gamma t}{2}-\frac{\Gamma\sin(2h_0 t)}{4h_0},
\end{equationS}
where the mean is squared \textit{only after} performing the average over measurement realizations. Since $\hat{x}$ and $\hat{p}$ are unbounded observables the distribution just keeps growing in time without properly reaching a steady state, but $\underline{\hat{\rho}}$ does start behaving like an infinite temperature state for observables that are concentrated near the origin of phase space.

\subsection{Designing the filter from experimentally accessible quantities}\label{sec:SSite:FilterExperimental}
The previous section illustrated that by adequately postprocessing the measurement record we can feasibly estimate properties of the conditional quantum state. Nevertheless, this required the analytical solution of the conditional dynamics to obtain the functional form of the filter given in Eq.~(\ref{eqn:SSite:MeanEstimator}). In reality this is not strictly necessary, and the functional form of the filter can in fact be extracted either from the measurement record $dI(s)$ (hence experimentally) or from the unconditional dynamics~\cite{Muller2009}. This can be shown by reframing the estimation of $\braket{\hat{x}}$ as the minimization of a cost functional
\begin{equation}\label{eqn:SSite:CostFunctionalGeneral}
    C=\lim_{T\to\infty}\frac{1}{T}\int_{-T/2}^{T/2}\overline{\big[\braket{\hat{x}(t)}-x_{\text{est}}(dI)\big]^2}dt,
\end{equation}
where for clarity we are now explicitly including the time dependence of $\braket{\hat{x}(t)}$. The estimator $x_{\text{est}}(dI)$ is a generic functional of the measurement record, and the cost functional involves a mean square average over measurement realizations and over time. The minimization is done with respect to all possible functionals $x_{\text{est}}$ of $dI(s)$ and the solution to this problem should provide the exact form of $\braket{\hat{x}}$ as a functional of the record $dI(s)$. In practice this is a very complicated problem (see Appendix~\ref{app:GeneralFilter} for a general although unusable solution), but we can always restrict the space of functionals to e.g. time-translationally invariant linear filters
\begin{equation}
    x_{\text{est}}^{\text{lin}}(dI)=\int_{-\infty}^t f(t-s)\,dI(s).
\end{equation}
In this case, the cost functional becomes
\begin{align}\begin{split}\label{eqn:SSite:CostFunctional}
    C(f)&=\lim_{T\to\infty}\frac{1}{T}\int_{-T/2}^{T/2}dt\bigg[\overline{\braket{\hat{x}(t)}^2}\\
    &-2\int_{-\infty}^{t}f(t-s)\overline{\braket{\hat{x}(t)}dI(s)}\\
    &+\int_{-\infty}^t\int_{-\infty}^t f(t-s)f(t-s')\overline{dI(s)dI(s')}\bigg],
\end{split}\end{align}
and should be minimized with respect to the function $f$. Note that the first line in the previous equation involves a nonlinear average but is independent of the minimization variables and will thus not contribute to the solution. Every other term in Eq.~(\ref{eqn:SSite:CostFunctional}) is at most linear in the conditional quantum state and so can in fact be calculated without postselection. Terms like $\overline{dI(s)dI(s')}$ [or in the general case $\overline{x_{\text{est}}(dI)x_{\text{est}}(dI)}$] can in principle be obtained by repeating the experiment many times and doing auto-correlations of the measurement record alone, while terms like $\overline{\braket{\hat{x}(t)}dI(s)}$ [or $\overline{\braket{\hat{x}(t)}x_{\text{est}}(dI)}$ in the general case] involve cross-correlations between the measurement record and direct measurements of $\hat{x}$ on the final quantum state, but do not require postselection. Because of this, these two types of contributions can be obtained directly from the unconditional dynamics. This is true for any general functional $x_{\text{est}}(dI)$, but we now illustrate this point for the case of a linear filter. 

Minimization of Eq.~(\ref{eqn:SSite:CostFunctional}) over $f$ leads to a linear equation that $f$ must satisfy:
\begin{equationS}\label{eqn:SSite:FilterSolution}
    &\int_0^\infty dt'\,f(t') S(t-t')=g_x(t),\hspace{1cm} t>0
\end{equationS}
where we have defined the record-record correlation function $S$ according to
\begin{equation}\label{eqn:SSite:Current-current0}
   S(t-t')dt dt'=\lim_{T\to\infty}\frac{1}{T}\int_{-T/2}^{T/2}ds\,\overline{dI(s-t)dI(s-t')},
\end{equation}
and the cross-correlator $g_O$ (for any operator $\hat{O}$):
\begin{equation}
    g_O(t)dt=\left[\lim_{T\to\infty}\frac{1}{T}\int_{-T/2}^{T/2}ds\,\overline{\braket{\hat{O}(s)}dI(s-t)}\right].
\end{equation}
To obtain $f$ we must invert the action of $S$. The Fourier transform of $S$ (proportional to the power spectrum of the measurement record~\cite{Jacobs_2014}) may have zeroes at some specific (complex) frequency values that will become the poles of $f$ and so will determine its behaviour at long times~\cite{Muller2009}. The structure of $g_O(t)$ will then add details about phase relations and timings. Both $g_O$ and $S$ can be calculated directly from the unconditional dynamics, and are naively given by~\cite{Wiseman1993,zoller1997quantumnoisequantumoptics}
\begin{equationS}\label{eqn:SSite:Current-current}
    S(t)&=\delta(t)+4\Gamma\, \mathrm{Re}\left\{\mathrm{Tr}\left[\hat{x}e^{\mathcal{L}|t|}(\hat{x}\rho_{\text{ss}})\right]\right\}\\
    g_O(t)&=2\sqrt{\Gamma}\,\mathrm{Re}\left\{\mathrm{Tr}\left[\hat{O}e^{\mathcal{L}t}(\hat{x}\rho_{\text{ss}})\right]\right\}
\end{equationS}
where the $\hat{x}$ operator appears because it is the observable that is being continuously measured, $\mathcal{L}(\hat{\rho})=-ih_0[\hat{a}^\dagger\hat{a},\hat{\rho}]+\Gamma(\hat{x}\hat{\rho}\hat{x}-\{\hat{x}^2,\hat{\rho}\}/2)$ is the unconditional evolution superoperator obtained from Eq.~(\ref{eqn:SSite:SSE}) by omitting the stochastic $\propto dW$ terms, and $\underline{\hat{\rho}_{ss}}$ is the unconditional steady state of the system defined by $\mathcal{L}(\underline{\hat{\rho}_{ss}})=0$.

Eq.~(\ref{eqn:SSite:Current-current}) are the generic forms of the record-record correlator $R$ and of the cross-correlator $g_O$, valid also for other systems once $\hat{x}$ is replaced by the appropriate measured observable, but the dynamics described by Eq.~(\ref{eqn:SSite:SSE}) is slightly pathological because it does not posses a proper unconditional steady state $\underline{\hat{\rho}_{ss}}$: the infinite temperature state in a boson system has on average $\braket{\hat{a}^\dagger\hat{a}}=\infty$. Such a state cannot be reached nor approached in finite time, so we need to be more careful about what we mean by the $T\to\infty$ limit in Eq.~(\ref{eqn:SSite:Current-current0}). In practice, the properties of $R$ lead to a differential equation for $f(s)$ (see Appendix~\ref{app:SSite:RecordRecord})
\begin{equation}\label{eqn:SSite:FilterEquation}
    \left[\left(\frac{d^2}{dt^2}+h_0^2\right)^2+4\Gamma^2h_0^2\right]f(t)=0.
\end{equation}
and the structure of $g_x$ imposes two conditions on $f$:
\begin{equationS}\label{eqn:SSite:FilterConditions}
    \int_{0}^{\infty}f(t)\cos(h_0 t)\,dt&=\frac{1}{2\sqrt{\Gamma}}\\
    \int_{0}^{\infty}f(t)\sin(h_0 t)\,dt&=0.
\end{equationS}
This is enough to determine $f(t)$. The differential equation indicates that the filter responds as an exponential $e^{\lambda t}$ with constants
\begin{equation}
    \lambda=\pm\left(\frac{1}{\tau}\pm i h_0\Gamma\tau\right),
\end{equation}
where $\tau=(2 \Gamma v_x^{\infty})^{-1}$ is a memory time, and the solutions that blow up exponentially can be omitted on physical grounds. Comparison with Eq.~(\ref{eqn:SSite:MeanEstimator}) shows that both the exponential decay constant and the oscillation frequency are correctly given by $e^{\lambda t}=e^{-t/\tau}e^{\pm i h_0\Gamma \tau t}$. Even without solving exactly for the filter we have already obtained valuable information about the conditional dynamics just by evaluating the response of the unconditional system. Further demanding Eq.~(\ref{eqn:SSite:FilterConditions}) to hold leads to Eq.~(\ref{eqn:SSite:MeanEstimator}).

Let us summarize and discuss in more depth the points made in this subsection. To begin with, the problem of finding the optimal filter [i.e. minimizing Eq.~(\ref{eqn:SSite:CostFunctional})] can be formulated directly in the unconditional dynamics. As discussed before, for a generic filter one needs to estimate $\overline{x_{\text{est}}(dI)x_{\text{est}}(dI)}$ and $\overline{x_{\text{est}}(dI)\braket{\hat{x}(t)}}$, both of which can be obtained either from the measurement record or from correlations between the measurement record and the final quantum state of the system. 

From an experimental standpoint, one could take advantage of this situation by running the conditional dynamics many times, building an ensemble of measurement records and then testing various possible filters, both linear and nonlinear. The choice between filters would then be determined by which one leads to a smaller $C$. For the purposes of comparison between different filters, the $\overline{\braket{\hat{x}(t)}^2}$ term in Eq.~(\ref{eqn:SSite:CostFunctionalGeneral}) is irrelevant because it is independent of the filter. Nevertheless, a global determination of the efficacy of the filter on its own (namely, how close $C$ is to $0$) is not possible precisely because $\overline{\braket{\hat{x}(t)}^2}$ is unknown and nonlinear in the conditional density matrix. It is still possible to use $\overline{x_{\text{est}}(dI)x_{\text{est}}(dI)}$ as a proxy for $\overline{\braket{\hat{x}(t)}^2}$ since they are equal for the optimal filter, but the resulting modified cost functional is no longer bounded below and so must be used with care.

For systems where a proper unconditional steady state $\underline{\hat{\rho}_{ss}}$ exists, the time integral in Eq.~(\ref{eqn:SSite:CostFunctional}) is unnecessary and the experimental determination of the filter could be done by probing at a fixed time that is long enough for the unconditional dynamics to reach $\underline{\hat{\rho}_{ss}}$, thus reducing the complexity of the procedure. The complexity of such a task can be reduced by constraining the filter structure based on physically reasonable principles such as locality in time (and space in spatially extended systems).

To finalize, we point out that a distinction needs to be made between the unconditional steady state $\underline{\hat{\rho}_{ss}}$ and the unconditional dynamics that leads to it, characterized by $\mathcal{L}$. While $\underline{\hat{\rho}_{ss}}$ is generally featureless, response functions [such as Eq.~(\ref{eqn:SSite:Current-current})] that probe the dynamical behaviour of perturbations away from $\underline{\hat{\rho}_{ss}}$ do carry important information about the conditional dynamics. In the case of Eq.~(\ref{eqn:SSite:Current-current}), the information about the continuously measured observable is encoded in the choice of operators within the correlation function ($\hat{x}$).

\subsection{Discussion}
One of the important features of this particular unitary+measurement dynamics is the emergence of a memory time. Indeed, a glance at Eq.~(\ref{eqn:SSite:MeanEstimator}) indicates that the mean $\braket{\hat{x}}$ depends only on the past values of the increments $dI(s)$ within a time $\tau=(2\Gamma v_x^{\infty})^{-1}$ of the observation time $T$. In general, any event outside of this time window does not meaningfully affect the quantum state at time $T$. Note that the existence of $\tau$ requires both unitary dynamics and measurements: if $\Gamma\to 0$ at fixed $h_0$, then $\tau$ diverges as $\tau\sim \Gamma^{-1}$, while for $h_0 \to 0$ at fixed $\Gamma$ we find that $\tau$ diverges as $\tau\sim (\Gamma h_0)^{-1/2}$. It is worth pointing out that the time $\tau$ also controls the equilibration of the covariances. Small perturbations about the steady state solutions of Eq.~(\ref{eqn:SSite:Covs}) will decay away with a time constant given by $\tau/2$. Thus, if $\tau=\infty$, the steady state covariances are unreachable in practice.
    
The existence of $\tau$ imposes restrictions on how the quantum state can depend on the measurement record, namely locality in time. However, the allowed dependence is even more restricted. Since the quantum state is entirely determined by means and covariances, the dependence on the record is entirely codified by the ``estimator" $x_{\text{est}}$ and the analogous quantity associated to $\braket{\hat{p}}$. Thus, the quantum state really depends only on two numbers that can be calculated from the measurement record. As long as these two numbers are the same, independently of what the individual increments $\{dI\}$ are, the quantum state will be the same.
\subsection{Recovering observables}\label{subsec:SSite:Postselection}
\begin{figure*}
    \centering
    \includegraphics[width=0.78\textwidth]{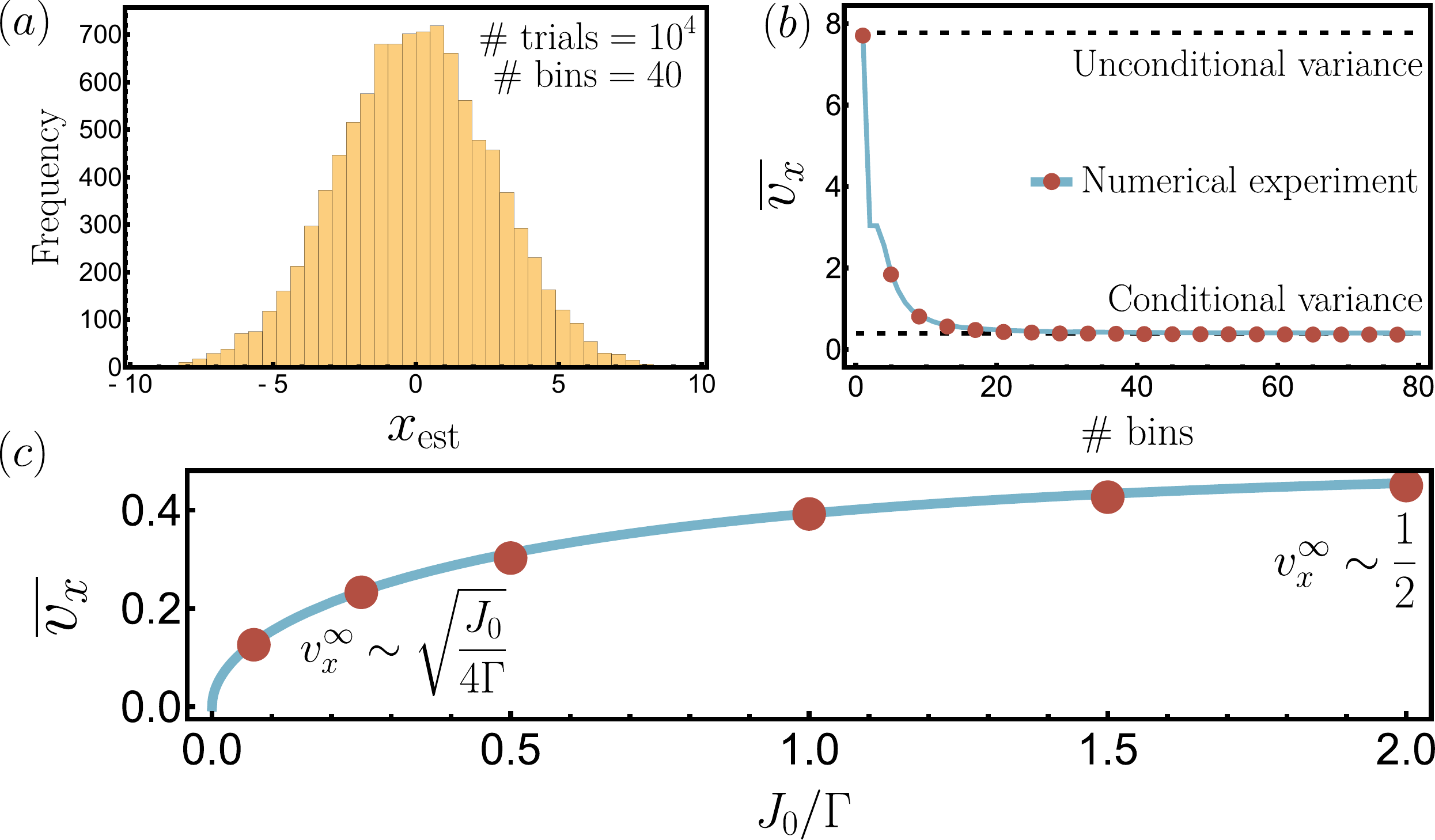}
    \caption{Results of the numerical implementation of the postselection procedure as outlined in section~\ref{subsec:SSite:Postselection} after evolving the initial state $\ket{0}+\ket{5}$ for $t=10\Gamma^{-1}$ with $\Gamma=h_0$ a number $N_{\text{trials}}=10^4$ of times. (a) Histogram of $x_{\text{est}}$ after binning with 20 bins. Each bin is associated to $\sim 100$ realizations of the experiment so statistics can be done with confidence. (b) Conditional variance after binning the measurement outcomes $x_{\text{meas}}$ according to $x_{\text{est}}$ as a function of the number of bins. One bin is the unconditional evolution. Error bars are smaller than the size of the red dots. (c) Dependence of $\overline{v_x}$ on $h_0/\Gamma$. Blue curve is Eq.~(\ref{eqn:SSite:VarianceSolution}) and red dots are the numerical results}
    \label{fig:SingleSite:Postselected}
\end{figure*}

We can test these ideas by recovering specific observables, which require even less information than the quantum state to be accessed, using only experimentally accesible quantities. Take $v_x$ for example. Obtaining the average $\overline{v_x}$ only requires $\braket{\hat{x}}$ to be estimated accurately, so we must build the postselected ensemble by lumping together measurement realizations that share the same value of $x_{\text{est}}$, within some tolerance. We thus put forward the following protocol designed to measure $\overline{v_x}$:
\begin{enumerate}
    \item Run the experiment once to obtain a single copy of the quantum state $\hat{\rho}$ and a single realization of the record $\{dI\}$.
    \item Sample a single value $x_{\text{meas}}$ by measuring the observable $\hat{x}$ in the quantum state $\hat{\rho}$, with associated probability distribution $P(x)=\braket{x|\hat{\rho}|x}$ ($\ket{x}$ is an eigenstate of $\hat{x}$).
    \item Construct $x_{\text{est}}$ using the specific $\{dI\}$ obtained in this iteration of the experiment. The output of the last two steps is the pair of numbers $(x_{\text{meas}},x_{\text{est}})$.
    \item Repeat the experiment $N_{\text{trial}}$ times to obtain $N_{\text{trial}}$ pairs $(x^r_{\text{meas}},x_{\text{est}}^r)$ for $r=1,... N_{\text{trial}}$.
    \item Bin the data according to the values of $x_{\text{est}}^r$. Since $\hat{x}$ is a continuous variable there will necessarily be some coarse-graining. The bin number and size should be chosen such that in most bins there are enough data points to do statistical averages with low sampling error.
    \item Calculate $\braket{\hat{x}}$ and $\braket{\hat{x}^2}$ in each bin using the experimentally measured $x_{\text{meas}}$ and doing statistics. This gives us $v_x^{\text{bin}}$. Because $v_x^{\text{bin}}$ depends on the measurement record only through $\braket{\hat{x}}\sim x_{\text{est}}$, this binning procedure allows us to access the quantum trajectory value of $v_x$. Referring back to Fig.~\ref{fig:SingleSite:Husimi}, each bin corresponds to a postselected distribution in which the $x_{\alpha}$ coordinate of the centroid has roughly the same value, so the only fluctuations come directly from the quantum state in the given trajectory. As long as the bin size is smaller than these quantum fluctuations we can be confident that we will recover the properties of the quantum trajectory.
    \item Finally, we average $v_x^{\text{bin}}$ over all bins to obtain $\overline{v_x}$. For this gaussian system $v_x^{\text{bin}}$ is bin independent and equal to $\overline{v_x}$, but this need not be the case for more general models.
\end{enumerate}
\begin{figure}
    \centering
    \includegraphics[width=0.48\textwidth]{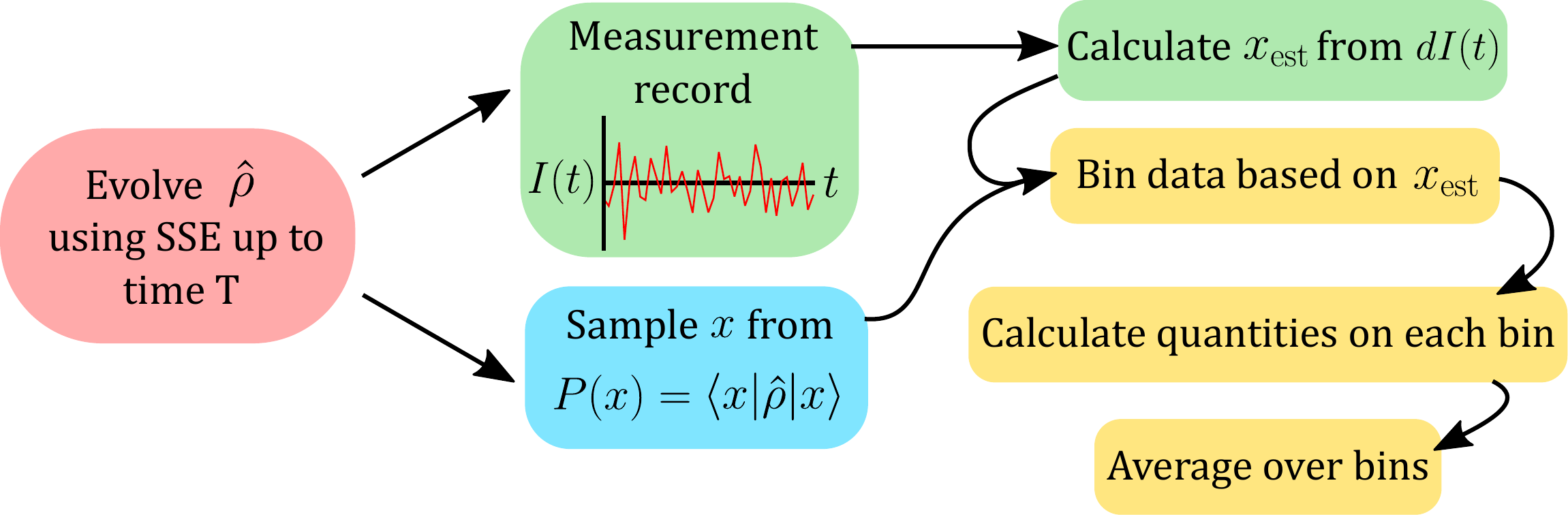}
    \caption{Flowchart of the protocol designed to recover the conditional variance on single quantum trajectories.}
    \label{fig:SingleSite:Flowchart}
\end{figure}
We depict a flow chart of this procedure in Fig.~\ref{fig:SingleSite:Flowchart}. We also test this protocol numerically and show the results in Fig.~\ref{fig:SingleSite:Postselected}. We used $N_{\text{trial}}=10^4$ realizations, beginning from the quantum state $\ket{0}+\ket{5}$ and evolving the system from $t=0$ to $t=10\Gamma^{-1}$ at $\Gamma=h_0$ using Eq.~(\ref{eqn:SSite:SSE}). We sample $x_{\text{meas}}$ directly from the quantum state at the end of each run and discard the state, leaving us only with $x_{\text{meas}}$ and the measurement record $dI$, from which we calculate $x_{\text{est}}$. Panel (a) presents a histogram of the $x_{\text{est}}$ values using $40$ bins of size $\delta x\approx 0.5$ each, and clearly shows that $x_{\text{est}}$ are distributed as a gaussian of mean $0$. Panel (b) shows the ensemble average $\overline{v_x}$ as a function of the number of bins. When the number of bins is $1$ (no postselection) we recover the unconditional variance, but we see that already at about $20$ bins the results have converged to the conditional result. This demonstrates that the binning procedure does not need to be fine-tuned, and a reasonable amount of coarse-graining is allowed. Finally, in panel (c) we repeat the protocol for different values of $h_0/\Gamma$ to show that we can recover the dependence of $\overline{v_x}$ on $h_0/\Gamma$ given by Eq.~(\ref{eqn:SSite:VarianceSolution}), which would naively be hidden behind the postselection barrier. 

In principle, the estimator $x_{\text{est}}$ provides the value of $\braket{\hat{x}}$ so that the variance $\overline{v_x}$ could also be determined directly from the measurement record. Nevertheless, this strategy might have a different degree of robustness against experimental imperfections, which is a question we leave for future work. Furthermore, the procedure put forward in this section gives us access to quantum state trajectories with fixed properties (fixed $x_{\text{est}}$ in this case), which is of relevance if these conditional quantum states are meant to be used for some task after preparation. This is the philosophy behind, e.g., the preparation of entangled states via quantum non-demolition measurements~\cite{Braginsky1996,Kuzmich2000,SchleierSmith2010,Cox2016,Hosten2016}.

\subsection{Non-gaussian states}\label{subsec:SSite:Nongaussian}
As promised, in this section we discuss the behaviour of non-gaussian states under the dynamics prescribed by Eq.~(\ref{eqn:SSite:SSE}). The analysis is more easily performed on quantum states rather than density matrices, so we work with~\cite{Jacobs2006}
\begin{equationS}
    d\ket{\psi}&=\left[-ih_0\hat{a}^\dagger\hat{a}-\frac{\Gamma}{2}(\hat{x}-\braket{\hat{x}})^2\right]\ket{\psi}\,dt\\
    &+\sqrt{\Gamma}\big(\,\hat{x}-\braket{\hat{x}}\big)\ket{\psi}dW
\end{equationS}
instead of Eq.~(\ref{eqn:SSite:SSE}), although they are equivalent. The general solution to this equation (in terms of $dI$) is given by (see Appendix~\ref{app:SSite:NonGaussian})
\begin{equation}\label{eqn:SSite:Non-GaussianEvolution}
    \ket{\psi}=e^{\hat{L}_1} e^{\hat{Q}t}e^{\hat{L}_2}\ket{\psi_0},
\end{equation}
where $\ket{\psi_0}$ is the initial state of the system and
\begin{equationS}
    \hat{Q}&=-ih_0\hat{a}^\dagger\hat{a}-\Gamma\hat{x}^2\\[5pt]
    \hat{L}_1&=\frac{\sqrt{\Gamma}}{2}\left(\hat{x}+\frac{h_0\tau\hat{p}}{F}\right)\int_0^te^{-F(t-s)}\,dI(s)\\
    \hat{L}_2&=\frac{\sqrt{\Gamma}}{2}\left(\hat{x}-\frac{h_0\tau\hat{p}}{F}\right)\int_0^te^{-Fs}\,dI(s),
\end{equationS}
where $F=\tau^{-1}+i\Gamma h_0\tau$ and $\tau=(2 \Gamma v_x^{\infty})^{-1}$ is the memory time. This decomposition has the following properties:
\begin{itemize}
    \item The first factor, $\hat{L}_2$, is linear in boson operators. Furthermore, the exponential factor in the integrand indicates that $\hat{L}_2$ only depends on the measurement record in the initial moments of the evolution. Because of this, it does not substantially modify the initial state at long times. 
    \item The second factor, $\hat{Q}$, is quadratic in boson operators and independent of the measurement record. At long times, it drives most states towards a fixed quantum state (see Appendix~\ref{app:SSite:NonGaussian}) that happens to be gaussian with zero mean and covariances given by the steady state solutions of Eq.~(\ref{eqn:SSite:Covs}).
    \item The last factor, $\hat{L}_1$, is again linear in boson operators and it modifies the means of the gaussian fixed point of $\hat{Q}$. In contrast to $\hat{L}_2$, the integral kernel has a convolution structure that enforces dependence on the measurement record only within a window $\tau$ of the observation time. This gives rise to the stochastic shift described by Eq.~(\ref{eqn:SSite:MeanEstimator}).
\end{itemize}

Our final remark before ending this section is that this decomposition implies that the late dynamics of any quantum state, even if it is initially non-gaussian, depends only on the two estimators introduced before, as was numerically verified in the gaussian case.

\section{Lattice model}\label{sec:Multisite}
All the discussion from the previous section will follow through for the lattice system, but now the spatial structure of the system will be reflected in the postselection procedure. We rewrite here the model for reference purposes
\begin{equationS}\label{eqn:Lattice:SSE}
    d\hat{\rho}&=-i\left[\hat{H},\hat{\rho}\right]\,dt+\Gamma\sum_i\left(\hat{x}_i\hat{\rho}\,\hat{x}_i-\frac{1}{2}\{\hat{x}_i^2,\hat{\rho}\}\right)\,dt\\
    &+\sqrt{\Gamma}\sum_i\Big(\hat{x}_i\hat{\rho}+\hat{\rho}\hat{x}_i-2\braket{\hat{x}_i}\hat{\rho}\Big)\,dW_i,
\end{equationS}
with Hamiltonian $\hat{H}=-J\sum_{\braket{ij}}\hat{a}_i^\dagger\hat{a}_j+J_0\sum_i \hat{a}^\dagger_i\hat{a}_i$ and measurement record $dI_i=2\sqrt{\Gamma}\braket{\hat{x}_i}\,dt+dW_i$. The gaussian equations of motion for second-order correlators take on a very similar form to Eq.~(\ref{eqn:SSite:Covs})
\begin{equationS}\label{eqn:MSite:Covs}
    \partial_t C^X&=h U^T+U h-4\Gamma (C^X)^2\\
    \partial_t C^P&=-h U -U^T h-4\Gamma U^T U+\Gamma\\
    \partial_t U&=h C^P-C^X h-4\Gamma C^X U,
\end{equationS}
where $(C^{X,P})_{ij}=C^{X,P}_{ij}$ are given in Eq.~(\ref{eqn:Def:CP}),
\begin{equationS}
    U_{ij}&=\frac{1}{2}\braket{\{\hat{x}_i,\hat{p}_j\}}-\braket{\hat{x}_i}\braket{\hat{p}_j}\\
    h_{ij}&=J_0\delta_{ij}-J\sum_{\mu}\delta_{i,j+\mu}
\end{equationS}
and $\mu$ indexes the spatial directions. Because of translational invariance, the steady state values of $C^{X,P}$ and $U$ can be obtained analytically by working in momentum space. More generally, for each momenta $q$, the evolution equation Eq.~(\ref{eqn:Lattice:SSE}) is a copy of the single site model Eq.~(\ref{eqn:SSite:SSE}) with $h_0$ replaced by $h_q=J_0-J\sum_{\mu}\cos(q_\mu)$, and all the momenta decouple from each other.

To achieve equilibration, the $h_q$ must be nonzero in the entire Brillouin zone. This can be enforced by requiring that $J_0>J_2$ (the case $J_0\to J_2^+$ is studied in Section~\ref{subsec:FilterFunctions}). As per Section~\ref{subsec:SSite:Nongaussian} this also implies that any initial non-gaussian state (with non-zero overlap with the vaccuum) will end up evolving into a gaussian state. Because of this, we will only consider gaussian states in what follows.

\begin{figure}
    \centering
    \includegraphics[width=0.48\textwidth]{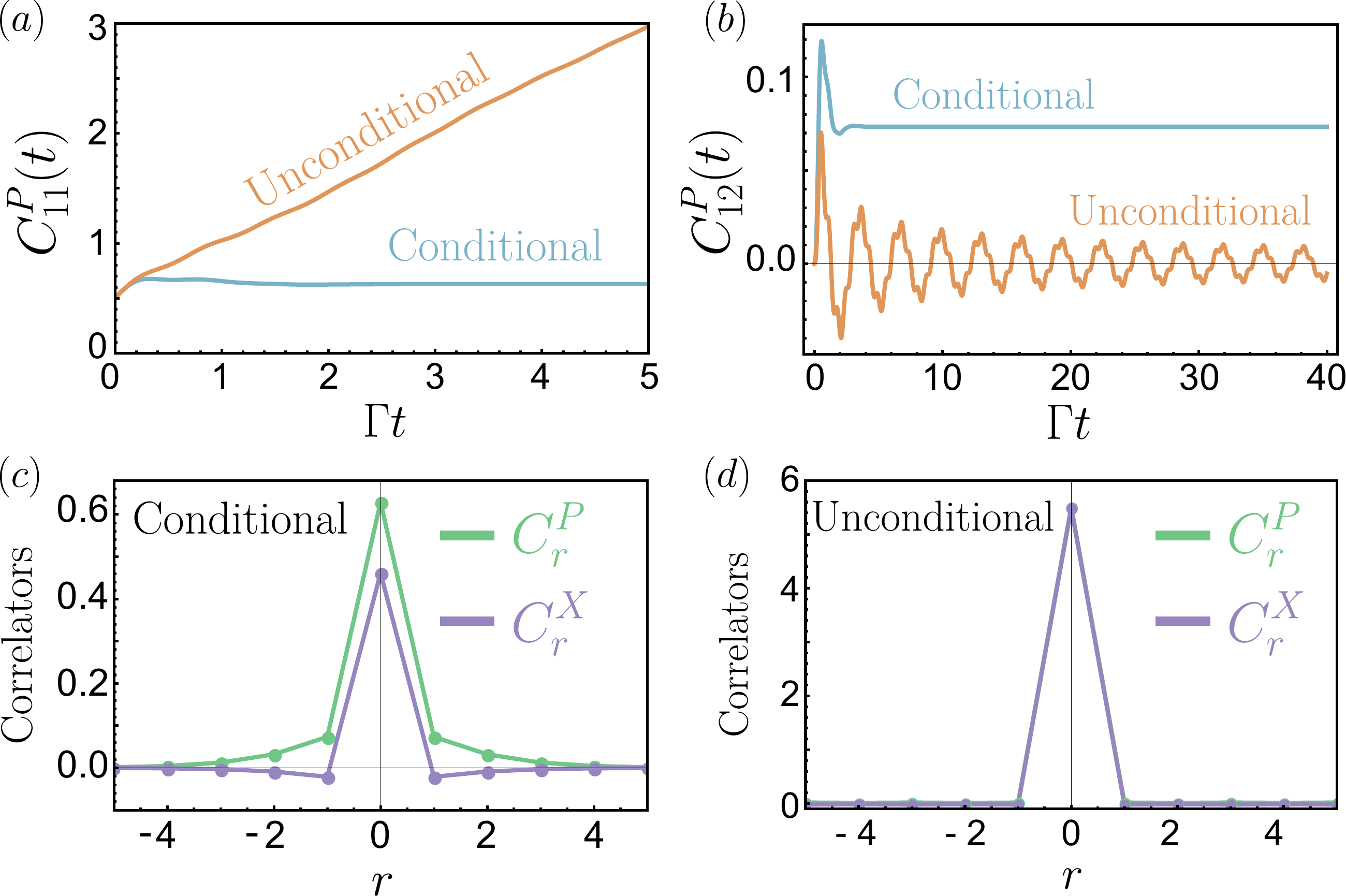}
    \caption{(a) Variance of $\hat{p}_1$ ($C^P_{11}$) as a function of time for an initial state with 0 bosons on the lattice. We show both the conditional and unconditional correlators. (b) Covariance of $\hat{p}_1$ and $\hat{p}_2$ ($C^{P}_{12}$) for the same initial state. (c) Equilibrium profile of the conditional correlators $C^{X,P}_r\equiv C^{X,P}_{i,i+r}$ as a function of separation between the operators. (d) Profile of the unconditional correlators at a time $t=10\Gamma ^{-1}$. $C^X$ and $C^P$ behave almost identically so the curves are on top of each other. Note the difference in vertical scale between (c) and (d).}
    \label{fig:MSite:Correlators}
\end{figure}

We use Eq.~(\ref{eqn:MSite:Covs}) to simulate the evolution of an initial state with 0 bosons in a one-dimensional $500$ site lattice. In Fig.~\ref{fig:MSite:Correlators}(a) we show the time evolution of $C^P_{11}$ (variance of $\hat{p}_1$) from $t=0$ to $t=5\Gamma^{-1}$ for $J_0=3\Gamma$, $J=\Gamma$ and observe fast equilibration. We also show the results for the unconditional evolution $\overline{\braket{\hat{p}_1^2}}$ (our initial state has $\overline{\braket{\hat{p}_1}}=0$ for all time), which keeps growing as in the single site case. Fig.~\ref{fig:MSite:Correlators}(b) does the same but for the two-site observable $C^P_{12}$ (covariance of $\hat{p}_1$ and $\hat{p}_2$). Again, the conditional evolution quickly equilibrates to a steady state, whereas the unconditional correlator decays slowly, and in an oscillatory fashion, towards $0$. Fig~\ref{fig:MSite:Correlators}(c) shows the conditional equilibrium profile of $C^{X,P}$ as a function of distance between lattice sites obtained after $t=10\Gamma ^{-1}$, when they have already reached their steady state. The correlators decay exponentially with a correlation length of the order of the lattice size. Finally, Fig.~\ref{fig:MSite:Correlators}(d) shows the same correlators but for the unconditional case, obtained at $t=10\Gamma^{-1}$, to demonstrate that their size and profile are very different from the conditional versions.

Our postselection procedure will be oriented towards recovering the equilibrium profile of the conditional correlators $C^X$ and $C^P$. To set this up, we need to investigate more closely the equations of motion for the means since these are the stochastic contributions that need to be estimated accurately. Again, they are formally very similar to the equations for the single site case, Eq.~(\ref{eqn:SSite:Means}), except that now we deal with matrices:
\begin{equation}\label{eqn:MSite:Means}
      d\begin{pmatrix}
        X\\ P
    \end{pmatrix}=\begin{pmatrix}
        -4\Gamma C^X&h\\-h-4\Gamma U^T&0
    \end{pmatrix}\begin{pmatrix}
        X\\ P
    \end{pmatrix}\,dt+2\sqrt{\Gamma}\begin{pmatrix}
        C^X\\
        U^T
    \end{pmatrix}\,dI,
\end{equation}
where $(X)_i=\braket{\hat{x_i}}$, $(P)_i=\braket{\hat{p}_i}$ and $(dI)_i=dI_i$ are vectors. At late times, when the covariance matrices have equilibrated, we can represent the solutions as spatio-temporal convolutions between the measurement record and filter functions. To be more concrete, we can write 
\begin{equationS}
(\hat{x}_i)_{\text{est}}(t)&=2\sqrt{\Gamma}\sum_j\int_{-\infty}^{t}K_x(i-j,t-s)dI_j(s)\\
(\hat{p}_i)_{\text{est}}(t)&=2\sqrt{\Gamma}\sum_j\int_{-\infty}^{t}K_p(i-j,t-s)dI_j(s)
\end{equationS}
where the filters $K_{x,p}(r,T)$ are given by
\begin{equationS}\label{eqn:MSite:Filters}
    K_x(r,T)&=\frac{1}{V}\sum_q \cos\left(\frac{h_q T}{2 v_q}\right) v_q\times e^{-2\Gamma v_q T} e^{iq r}\\
    K_p(r,T)&=\frac{1}{V}\sum_q \left[2\cos\left(\frac{h_q T}{2 v_q}\right) u_q-\sin\left(\frac{h_q T}{2 v_q}\right)\right]\\
    &\hspace{1.5cm}\times e^{-2\Gamma v_q T}e^{iq r},  
\end{equationS}
and $v_q, u_q, w_q$ are the multisite version of the steady-state covariances $v_x, u, v_p$, respectively [compare with Eq.~(\ref{eqn:SSite:VarianceSolution})]:
\begin{equationS}\label{eqn:MSite:VarianceSolution}
    v_q&=\frac{\sqrt{\frac{h_q}{8\Gamma}}}{\left(\frac{h_q}{4\Gamma}+\sqrt{\left(\frac{h_q}{4\Gamma}\right)^2+\frac{1}{4}}\right)^{1/2}}\\
    u_q&=\frac{1/4}{\frac{h_q}{4\Gamma}+\sqrt{\left(\frac{h_q}{4\Gamma}\right)^2+\frac{1}{4}}}\\
    w_q&=\sqrt{\frac{2\Gamma}{h_q}}\frac{\sqrt{\left(\frac{h_q}{4\Gamma}\right)^2+1}}{\left(\frac{h_q}{4\Gamma}+\sqrt{\left(\frac{h_q}{4\Gamma}\right)^2+\frac{1}{4}}\right)^{1/2}}.
\end{equationS} 
From these we can get the real space steady-state two-point correlators:
\begin{equation}\label{eqn:MSite:CorrelatorProfiles}
    (C^X_{jk},C^P_{jk})=\sum_q  e^{iq\cdot (j-k)} (v_q,w_q).
\end{equation}
Recovering the full quantum state requires the correct estimation of all $\braket{\hat{x}_i}$ and $\braket{\hat{p}_i}$. We thus run again into a postselection barrier where we would need to reproduce the results of an extensive number of measurements to build an ensemble of identical quantum states. Note, however, that we need one number per lattice site, not per time step, so this extensivity does not extend into the time variable. Furthermore, local correlators should only depend on the properties of the quantum state near the probed region, so the postselection overhead should still be substantially reduced if we want to recover $C^{X,P}$. Before discussing these solutions in more detail, let us in fact show that we can recover the equilibrium profiles of the conditional $C^{X,P}$.
\subsection{Post-selection}\label{subsec:MSite:Postselection}
\begin{figure}
    \centering
    \includegraphics[width=0.48\textwidth]{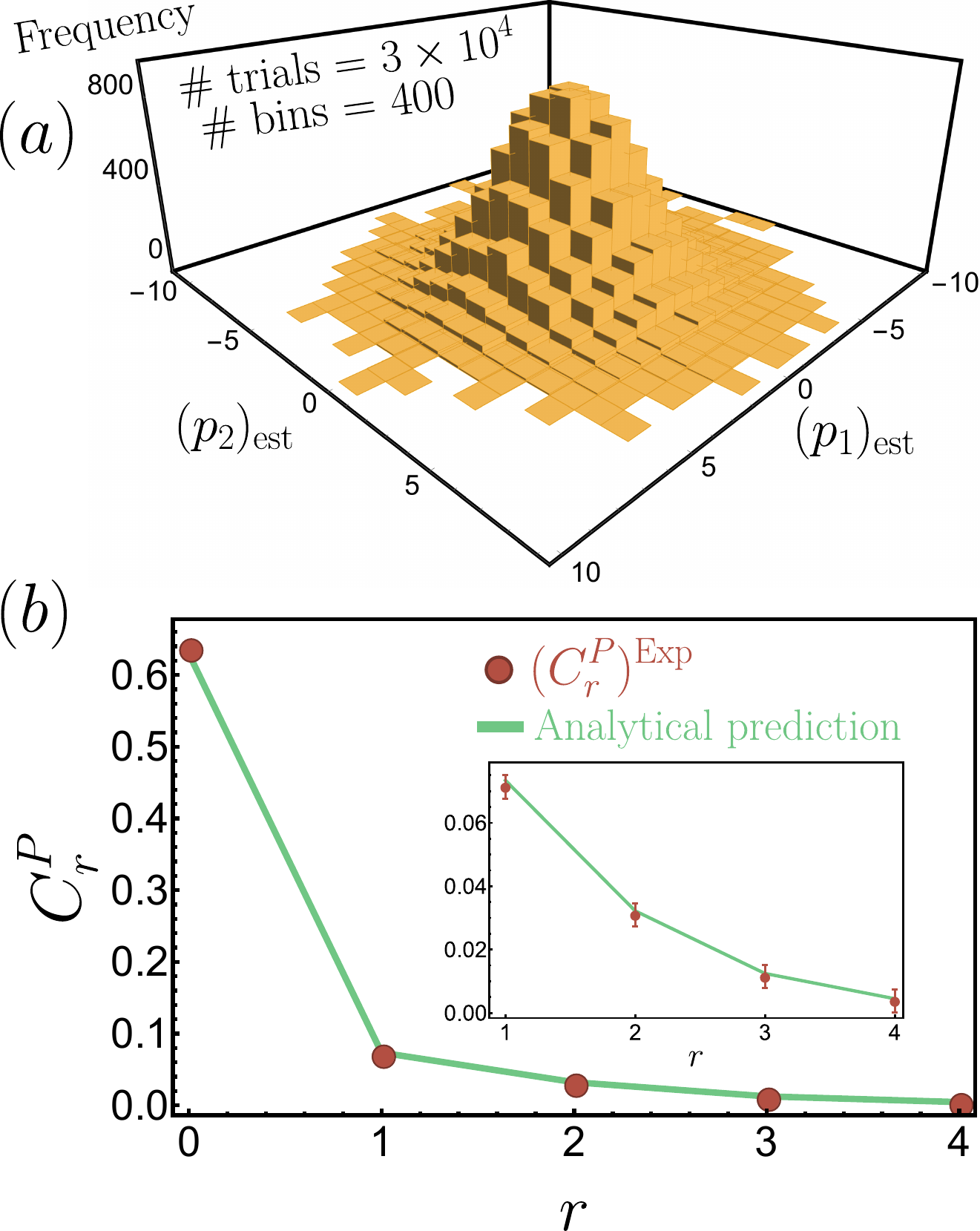}
    \caption{(a) Binning used to recover $C_{12}^P$ from experimentally accesible data. This time the binning is two-dimensional. (b) Comparison between the ``experimentally recovered" equilibrium profile of $C^P(r)\equiv C_{1,1+r}^P$ (red dots) and the analytical expectation (green line). Inset is a close up of the correlators for $r\geq 1$, showing the size of the statistical error bars.}
    \label{fig:MSite:Postselection}
\end{figure}

We repeat here the procedure put forward in Section~\ref{subsec:SSite:Postselection} but we will focus on recovering $C^P$ because it is numerically larger than $C^X$ (the measurements reduce $\hat{x}$ variances and consequently enlarge $\hat{p}$ variances):
\begin{enumerate}
    \item Run the experiment once to obtain a single copy of the quantum state $\hat{\rho}$ and a single realization of the record $\{dI_i\}$.
    \item Sample a single value $(p_i)_{\text{meas}}$ for each lattice site $i$ by measuring all the observables $\hat{p}_i$ in the quantum state $\hat{\rho}$. They all commute, so this is allowed.
    \item Construct $(p_i)_{\text{est}}$ from the specific $\{dI_i\}$ obtained in this iteration of the experiment. The output of this single run of the experiment is a pair of numbers for each lattice site $\{(p_i)_{\text{meas}},(p_i)_{\text{est}}\}$.
    \item Repeat the experiment $N_{\text{trial}}$ times to obtain $N_{\text{trial}}$ collections of $\{(p_i)^r_{\text{meas}},(p_i)^r_{\text{est}}\}$ for $r=1,... N_{\text{trial}}$.
    \item If we want to recover $C^P_{1,2}$, for example, we then bin the data according to the values of $(p_1)_{\text{est}}^r$ and $(p_2)_{\text{est}}^r$. This results in a two dimensional binning procedure (except for the onsite variance), which is slightly more intensive than the procedure for a single site, but not unmanageable.
    \item Calculate $\braket{\hat{p}_1}$, $\braket{\hat{p}_2}$ and $\braket{\hat{p}_1\hat{p}_2}$ in each bin using the experimentally measured $(p_1)^r_{\text{meas}}$ and $(p_2)^r_{\text{meas}}$ and doing statistics. This gives us $(C^P_{1,2})^{\text{bin}}$ .
    \item Finally, we average $(C^P_{1,2})^{\text{bin}}$ over all bins to obtain $(C^P_{1,2})^{\text{exp}}$. 
    \item If we now want to calculate $(C^P_{1,3})^{\text{exp}}$ we use this same data set, but now we bin the result according to $(p_1)_{\text{est}}^r$ and $(p_3)_{\text{est}}^r$. By doing different binnings, we can then obtain all the $(C_{ij}^P)^{\text{exp}}$ from the same data set.
\end{enumerate}
We implement this protocol numerically for $J_0=3\Gamma$, $J=\Gamma$ and run the experiment from $t=0$ to $t=10\Gamma^{-1}$ beginning from the vaccuum state ($\ket{0}$ in each lattice site). To evolve the ``quantum state" we use the gaussian evolution equations for means and covariances with appropriate initial conditions ($X=P=0$, $U=0$, $C^X=C^P=1/2$) and use this information to sample $(p_i)_{\text{meas}}$ from the underlying probability distribution at the end of the run. We then discard the means and covariances since this is information unavailable in an actual experiment, and are left with $(p_i)_{\text{meas}}$ and the measurement record $\{dI_i\}$, from which we calculate $(p_i)_{\text{est}}$ for each run. We repeat this a number $N_{\text{trials}}=3\times 10^4$ of times. We show the results in Fig.~\ref{fig:MSite:Postselection}. Panel (a) depicts the binning procedure used to calculate $C_{12}$. Panel (b) shows the equilibrium profile of $C^P(r)\equiv C^P_{1,1+r}$ obtained using the binning procedure and contrasts it against the analytically calculable profile [Eq.~(\ref{eqn:MSite:CorrelatorProfiles})], showing that the protocol indeed recovers the correlator that was ``hidden" behind postselection.

\subsection{Filter functions}\label{subsec:FilterFunctions}
\begin{figure}
    \centering
    \includegraphics[width=0.48\textwidth]{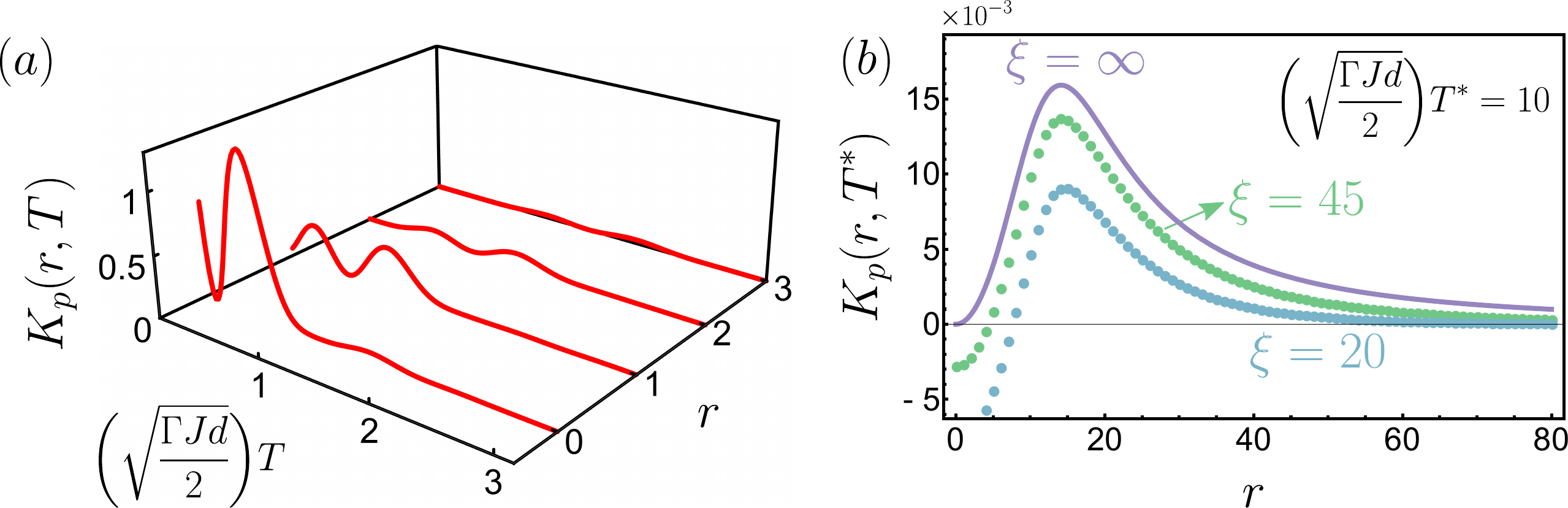}
    \caption{(a) Filter $K_p(r,T)$ in $d=1$ as a function of $r$ and $T$ for $J_0=3\Gamma$, $J=\Gamma$. (b) Filter $K_p(r,T)$ obtained from Eq.~(\ref{eqn:MSite:Filters}) in $d=1$ as function of $r$ for fixed $\Big(\sqrt{\Gamma Jd/2}\Big)T^*=10$ and different correlation lengths $\xi=20, 45,\infty$, corresponding, respectively, to $J_0=2.0025\Gamma,2.0005\Gamma, 2\Gamma$.}
    \label{fig:MSite:Filter}
\end{figure}

Let us now study a bit more in detail the filter function $K_p(r,T)$ defined in Eq.~(\ref{eqn:MSite:Filters}) because its structure determines how the measurement record must be postprocessed. Fig.~\ref{fig:MSite:Filter}(a) shows $K_p(t,T)$ for a $1d$ lattice system at $J_0=3\Gamma$, $J=\Gamma$, and illustrates that $K_p$ is concentrated near $r,T=0$, with a decay length of the size of the lattice spacing and an overall memory time $\tau$, determined by the smallest $h(q)$. Because of the convolution structure in Eq.~(\ref{eqn:MSite:Means}), estimation of $\braket{\hat{p}_j}(t)$ only requires knowledge of the measurement records within a few lattice spacings of $j$ and up to a time $\tau$ into the past. For this specific set of parameters, all the relevant features of the postselection procedure occur at the lattice scale.

\begin{figure}
    \centering
    \includegraphics[width=0.48\textwidth]{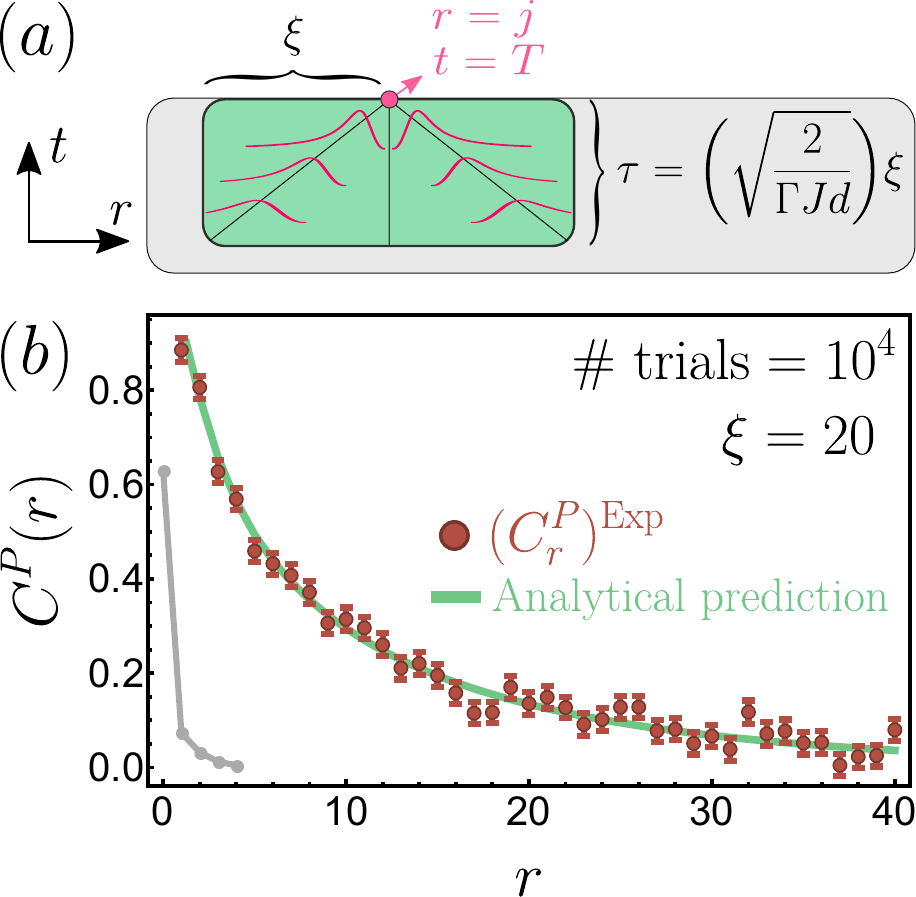}
    \caption{(a) Schematic depiction of the correlation volume (green) with spatial extent $\sim \xi$ and duration in time equal to the memory time $\tau=\xi\sqrt{2/(\Gamma Jd)}$. The maxima of the filter $K_p$ are localized along the diagonals, indicating ballistic propagation with velocity $\sqrt{\Gamma Jd/2}$. (b) $C^P(r)$ recovered from the postselection procedure of Section~\ref{subsec:MSite:Postselection} for $J_0=2.0025 \Gamma$ and $J=\Gamma$,  with correlation length $\xi=20$. Gray line is the correlator from Fig.~\ref{fig:MSite:Postselection}(b).}
    \label{fig:MSite:ContinuumCorrelator}
\end{figure}

By tuning $J_0$ , it is possible to increase the decay length and memory time, leading to features at much longer spatial/time scales that can be analyzed within a continuum description. In this regime, the $q$ integral that defines $K_p(r,T)$ [Eq.~(\ref{eqn:MSite:Filters})] is dominated by the small $q$ region. We can thus approximate $u_q\approx 1/2$,
\begin{equationS}
    h_q&\approx \frac{J}{2}(q^2+\xi^{-2})\\
    v_q&\approx \sqrt{\frac{J(q^2+\xi^{-2})}{4\Gamma}},
\end{equationS}
where $q^2=\sum_\mu q_{\mu}^2$ and we have defined a correlation length 
\begin{equation}
    \xi=\sqrt{\frac{J}{2(J_0-dJ)}}
\end{equation} that is assumed to be $\xi\gg 1$. Under these conditions, the filter becomes
\begin{equationS}
    &K_p(r,T)\approx \\
    &\sqrt{2}\mathrm{Re}\,\left[\int\frac{d^dq}{(2\pi)^d}e^{-(1+i)T\sqrt{\frac{J\Gamma d}{2}}\sqrt{q^2+\xi^{-2}}+iqr-\frac{i\pi}{4}}\right]
\end{equationS}
For fixed $\tau$ and large $r$, this decays exponentially as $e^{-r/\xi}$. For fixed $r$ and large $T$ the exponential decay is $e^{-\sqrt{(J\Gamma d/2)}T/\xi}$. When $\xi$ is large this defines a correlation volume in spacetime spanning multiple lattice spacings and lattice times. Beyond the borders of this volume the filter is essentially zero, indicating that the measurement record on those regions is irrelevant. Within that volume, the filter has a power-law form, as shown in Fig.~\ref{fig:MSite:Filter}(b) for increasing $\xi$ in $d=1$. In the $d=1$ case, for example, $K_p$ takes the form 
\begin{align}\begin{split}\label{eqn:MSite:FilterPowerLaw}
    K_p(r,T)\approx \frac{1}{2\pi}\Bigg[&\frac{r}{\left(\frac{J\Gamma d}{2}\right)T^2+\left(r-\sqrt{\frac{J\Gamma d}{2}}T\right)^2}\\
    &-\frac{r}{\left(\frac{J\Gamma d}{2}\right)T^2+\left(r+\sqrt{\frac{J\Gamma d}{2}}T\right)^2}\Bigg].
\end{split}\end{align}
For fixed $T$, this expression has maxima centered at $r\approx \pm T(J\Gamma d/2)^{1/2}$, reflecting ballistic spreading of information, and these maxima have widths $\delta r\approx T\sqrt{J\Gamma d/2}$. We summarize graphically these statements in Fig.~\ref{fig:MSite:ContinuumCorrelator}(a). For the two-point correlators near lattice sites $j$ and $k$, we need to carve correlation volumes around these regions, but the measurement record in the regions in between is irrelevant. For completeness, we repeat the numerical experiment of Section~\ref{subsec:MSite:Postselection} for parameters that are closer to a continuum description ($\xi=20$) and show the results in Fig.~\ref{fig:MSite:ContinuumCorrelator}(b). We see that we can still recover the decaying spatial profile of $C^P(r)$.

When $J_0<2dJ_2$, then $h(q)=J_0-J_2\sum_\mu \cos(q_\mu)$ will have zeroes at finite values of $q=q_{\text{zero}}$, and the long time dynamics will be dominated by long wavelength fluctuations on top of the rapid variations at these $q_{\text{zero}}$ values. The equilibration timescale for these fluctuations will be longer the longer their wavelength is. As a consequence, at any fixed time the shorter wavelengths will have gaussian correlations that are initial state independent (as in Section~\ref{subsec:SSite:Nongaussian}), whereas longer wavelengths will still retain information about the initial conditions. Entanglement entropies in this regime are expected to grow faster than the area of the subregion of interest~\cite{Minoguchi2022}.

\subsection{Application to single samples}
Based on the information we have gained about postselection in this lattice system, let us close the theoretical discussion by trying to frame the measurement-induced behaviours of this section as characteristics of a phase of matter. We follow the considerations put forward in Ref.~\cite{Friedman2023}:

\begin{enumerate}
    \item We begin with a \textit{single sample} of the system, of size $L^d$, and run the joint measurement+unitary dynamics \textit{only once}. We have the following resources at our disposal: the nature of the measurements that were done on the system ($\hat{x}$ in Section~\ref{sec:Multisite}), the measurement record (one function of time per lattice site), and a single copy of the quantum state.
    \item We now \textit{assume} that we can replace averages over different measurement realizations by averages over different spatial regions. Although we do not present a formal proof, we justify this assumption by arguing that spatial regions separated by more than few correlations lengths $\xi$ are uncorrelated in practical terms. This grants us access to unconditional correlators, as well as record-record and system-record correlations (as defined in Section~\ref{sec:SSite:FilterExperimental}). Even if $\xi$ is unknown, one could repeat the averaging with another set of spatial regions that are further separated from each other and expect convergence as this separation increases. If $\xi=\infty$, convergence might still be attainable, but will require larger separations.
    \item Using unconditional, record-record, and system-record correlators, we can employ the machinery of Section~\ref{sec:SSite:FilterExperimental} to design appropriate filters. 
    \item With the use of the filters, we can now extract information about this single quantum trajectory and perform nonlinear averages, once again using spatial averaging as a proxy for averages over measurement realizations. Note also that once the filters are obtained, nonlinear averages such as 
    \begin{equation}
        \overline{\braket{\hat{p}_1}\braket{\hat{p}_2}}
    \end{equation}
     can be replaced by estimators
     \begin{equation}
         \overline{(p_1)_{\text{est}}(p_2)_{\text{est}}},
     \end{equation}
     which require only the measurement record. Although these are equivalent in an ideal scenario, they might not be equally robust in the presence of experimental imperfections.
\end{enumerate}
In more complicated systems, steps (1), (2) and (4) should be unmodified, but the determination of the filters [step (3)] will be substantially more complex.

\section{Experimental implementations}\label{sec:ExpImp}
In this section we will describe possible experimental implementations of the models studied in this paper. As we will see, measurements of $\hat{x}$ require some degree of active control because they create bosonic excitations. The single site model is ubiquitous in cavity optomechanics, but we provide two different realizations in alternate platforms that possess interesting non-linear generalizations: cavity-QED and circuit-QED. The circuit-QED proposal will then be used as a building block for the lattice model.

\begin{figure}
    \centering
    \includegraphics[width=0.98\linewidth]{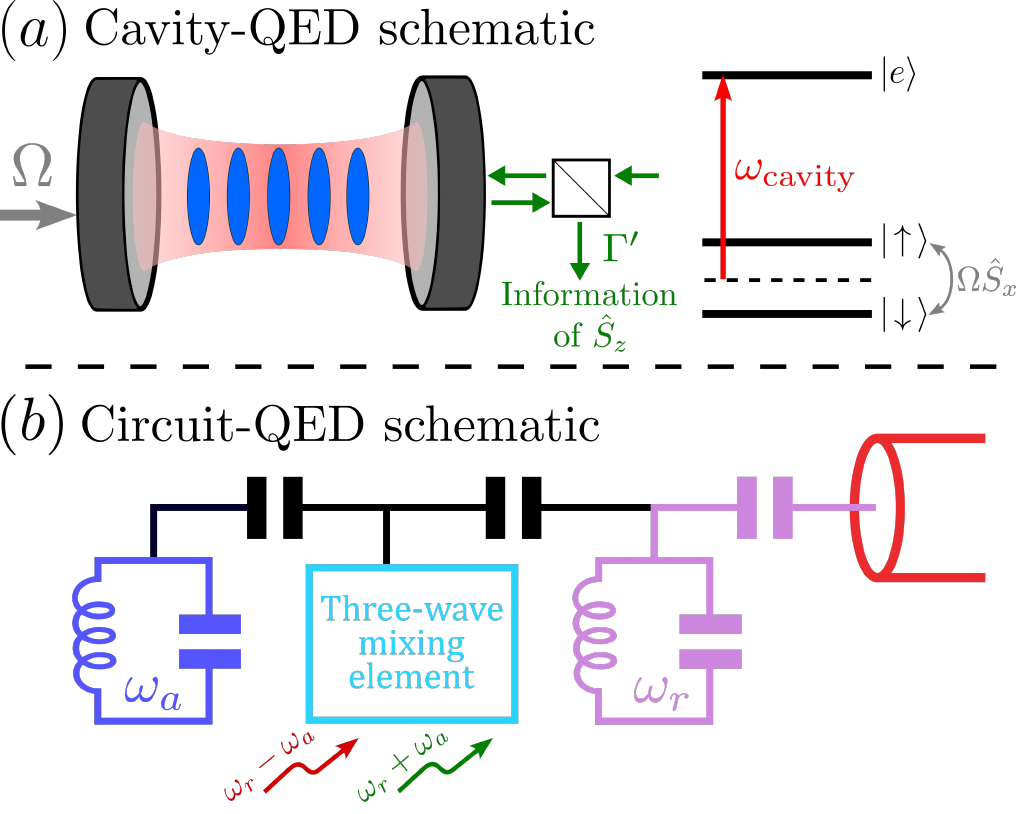}
    \caption{Schematics of the proposed implementations. (a) Cavity-QED system, where the atoms (blue pancakes) inside a cavity are driven ($\Omega\hat{S}_x$) and measured by probing the reflection of a laser tone sent towards the cavity. (b) Circuit-QED implementation, where two cavities ($a$ and $r$) are coupled via a three-wave mixing element, while the readount $r$ cavity is coupled to a transmission line, which carries information about the $a$ cavity.} 
    \label{fig:ImplementationSchematics}
\end{figure}

The first setup is based on cavity-QED implementations of quantum non-demolition measurements of atomic inversion~\cite{SchleierSmith2010,Cox2016,Hosten2016}. For definiteness, we consider an ensemble of $^{87}$Rb atoms inside a high-finesse optical cavity [see Fig.~\ref{fig:ImplementationSchematics}(a)]. Each atom can be approximated to be a two-level system, with  two hyperfine states defining the two-level manifold, $\ket{\downarrow}$ and $\ket{\uparrow}$ (the choice of hyperfine states depends on the specific experimental scheme). Taken together, the atoms give rise to collective spin operators 
\begin{equationS}
    \hat{S}_z&=\frac{1}{2}\sum_{k=1}^N (\ketbra{\uparrow}{\uparrow}_k-\ketbra{\downarrow}{\downarrow}_k)\\
    \hat{S}_x&=\frac{1}{2}\sum_{k=1}^N(\ketbra{\uparrow}{\downarrow}_k+\ketbra{\downarrow}{\uparrow}_k),
\end{equationS}
where $k$ indexes the $N$ atoms. The interaction between the atoms and the cavity [via a third level $\ket{e}$, see Fig.~\ref{fig:ImplementationSchematics}(a)] renormalizes the resonance frequency of the cavity by an amount that depends on the atomic state in the $\ket{\uparrow},\ket{\downarrow}$ manifold. Because of this, photons that escape from the cavity carry with them information about the atomic degrees of freedom that can then be accessed through measurements.

To realize the single-site model we send a laser tone, resonant with the bare cavity (i.e. when there are no atoms), along the cavity axis and probe the reflected light using homodyne detection. The reflected light will be phase shifted by an amount that depends on the renormalized cavity frequency and hence on the state of the atoms. Measuring this phase shift implements a continuous measurement of the atomic variable $\hat{S}_z$. Furthermore, we drive coherently the $\ket{\uparrow}\to\ket{\downarrow}$ transition using a microwave drive or two-photon Raman transitions, leading to a Hamiltonian term $\Omega\hat{S}_x$, where $\Omega$ is a Rabi frequency. The resulting evolution equation for the system, in the rotating frame of the drive, is 
\begin{equationS}\label{eqn:ExpImp:SpinModel}
    d\hat{\rho}=&-i\left[\Omega\hat{S}_x,\hat{\rho}\right]\,dt+\Gamma'\left(\hat{S}_z\hat{\rho}\,\hat{S}_z-\frac{1}{2}\{\hat{S}^2_z,\hat{\rho}\}\right)\,dt\\
    &+\sqrt{\Gamma'}\Big(\hat{S}_z\hat{\rho}+\hat{\rho}\hat{S}_z-2\braket{\hat{S}_z}\hat{\rho}\Big)\,dW,
\end{equationS}
and the result of the measurements is a homodyne current $dI=2\sqrt{\Gamma'}\braket{\hat{S}_z}dt+dW$. In this equation, the parameter $\Gamma'$ is derived and depends on properties of the cavity, of the laser used to probe the atoms, and the atom-light interaction strength~\cite{Davies2016}. If the initial state of the atoms is chosen to be along the $+x$ axis in the Bloch sphere, i.e. $\propto(\ket{\uparrow}+\ket{\downarrow})^{\otimes N}$, we can use the Holstein-Primakoff approximation to express the spin variables in terms of an auxiliary boson~\cite{Holstein1940}:
\begin{equationS}\label{eqn:ExpImp:HolsteinPrimakoff}
    \hat{S}_z&\approx \sqrt{\frac{N}{4}}(\hat{a}+\hat{a}^\dagger)=\sqrt{\frac{N}{2}}\hat{x}\\
    \hat{S}_x&\approx\frac{N}{2}-\hat{a}^\dagger\hat{a}.
\end{equationS}
This leads to Eq.~(\ref{eqn:SSite:SSE}) with $h_0=-\Omega$ and $\Gamma=\Gamma' N/2$. After evolving the system with Eq.~(\ref{eqn:ExpImp:SpinModel}) for some amount time, the final measurement of $\hat{x}\propto \hat{S}_z$ can be done by switching off the microwave/Raman drives [setting $\Omega\to 0$ in Eq.~(\ref{eqn:ExpImp:SpinModel})] and performing a QND measurement of $\hat{S}_z$. In this way, one could test the relation between the final, measured value of $\hat{x}$ and the estimate provided by the homodyne current before the drive is turned off. We could also repeat the procedure of Section~\ref{subsec:SSite:Postselection} to recover the variance, although this might be technically more challenging.

Relevant imperfections include finite detection efficiency, incoherent processes induced by spontaneous emission from $\ket{e}$ (virtually populated), and finite $N$ effects. Finite detection efficiencies can be incorporated naturally into Eq.~(\ref{eqn:SSite:SSE}) and do not qualitatively modify the results of Section~\ref{sec:SingleSite} but do change details, such as the exact form of the filter Eq.~(\ref{eqn:SSite:MeanEstimator}) and the absolute size of the $\hat{x}$ variance~\cite{Minoguchi2022}. Incoherent processes modify the long time behaviour of the system but they happen slowly and thus set a finite time window within which the pure drive+measurement dynamics can be observed. Finally, the large $N$ approximation [Eq.~(\ref{eqn:ExpImp:HolsteinPrimakoff})] will eventually break down because $\braket{\hat{x}}$ explores regions of phase space farther away from the origin as time increases (see Fig.~\ref{fig:SingleSite:Husimi}). The ensuing spin dynamics, although interesting, is something that we leave for future work.

The second setup requires the use of two superconducting microwave resonators, and is based on the proposals for cavity-based axion detectors of Refs.~\cite{Wurtz2021,Jiang2023}. We consider one mode in each cavity (with annihilation operators $\hat{a}$ and $\hat{r}$), that interact via simultaneous state-swapping and two-mode squeezing processes, generated by driving a three-wave mixing device (such as a SNAIL element~\cite{Frattini2017} or a Josephson Ring Modulator~\cite{Wurtz2021,Jiang2023}). The resulting Hamiltonian is
\begin{equationS}
    \hat{H}_{\text{modes}}&=\omega_a\hat{a}^\dagger\hat{a}+\omega_r\hat{r}^\dagger\hat{r}+\underbrace{g( e^{-i\omega_\Delta t}\hat{a}\hat{r}^\dagger+\text{h.c.})}_{\text{State swapping}}\\
    &+\underbrace{g'(e^{i\omega_{\Sigma}t}\hat{a}^\dagger\hat{r}^\dagger)+\text{h.c.}}_{\text{Two-mode squeezing}},
\end{equationS}
where $\omega_a$ ($\omega_r$) are the resonance frequencies of the system (readout) mode [see Fig.~\ref{fig:ImplementationSchematics}(b)]. The coupling strengths $g$, $g'$ and the modulation frequencies $\omega_\Delta\approx \omega_r-\omega_a$, $\omega_\Sigma\approx \omega_a+\omega_r$ are controlled by the microwave drives that are sent to the three-wave mixer. Choosing the drives so that $g=g'$, $\omega_\Delta=\omega_r-\omega_a+h_0$ and $\omega_\sigma=\omega_r+\omega_a-h_0$, leads, in an appropriate rotating frame, to the effective Hamiltonian
\begin{equationS}
    \hat{H}'_{\text{modes}}&=h_0\hat{a}^\dagger\hat{a}+g(\hat{a}+\hat{a}^\dagger)(\hat{r}+\hat{r}^\dagger)
\end{equationS}
To realize the measurements, we couple the readout mode to a transmission line (with strength $\kappa$) and do homodyne detection of the outgoing microwaves (in an appropriately chosen quadrature, referenced to the microwave drives). The coupled mode system is then described by the stochastic equation~\cite{Wiseman_Milburn_2009}
\begin{equationS}
    d\hat{\rho}&=-i[\hat{H}'_{\text{modes}},\hat{\rho}]\,dt+\kappa\left(\hat{r}\hat{\rho}\hat{r}^\dagger-\frac{\{\hat{r}^\dagger\hat{r},\hat{\rho}\}}{2}\right)\,dt\\
    &+\sqrt{\kappa}\left(\hat{r}\hat{\rho}+\hat{\rho}\hat{r}^\dagger-\braket{\hat{r}+\hat{r}^\dagger}\hat{\rho}\right)\,dW.
\end{equationS}
When $\kappa\gg h_0,g$, the readout mode can be adiabatically eliminated, leading to Eq.~(\ref{eqn:SSite:SSE}) with $\Gamma=8g^2/\kappa$. In other words, we are fixing the hierarchy of scales $\kappa\gg g\gg \Gamma=8g^2/\kappa\sim h_0$, to achieve the Purcell limit. This type of configuration is routinely accessed in circuit-QED setups.

For the lattice model, we would couple many of these two resonator systems together. There would be one transmission line per site, attached to the local readout cavity and carrying the local measurement results. The $\hat{a}$ cavities can then be capacitively coupled to allow for the transfer of photons between them. Although cross-talk between elements may be a relevant complication, simulation of similar lattice models in circuit-QED, even including interactions, has already been demonstrated in the past~\cite{Roushan2017,Ma2019,Roberts2023thesis}.

\section{Summary, conclusions and outlook}

In this paper we have studied the postselection problem in a system of bosons subjected to both unitary dynamics and continuous measurements. We first analyzed the case of a single site, which illustrates in a simple setting many important features of postselection, and then considered the extension to many sites to investigate the interplay between these features and the spatial structure of extended systems.

More concretely, we showed that to recover observables that are nonlinear in the conditional density matrix it is not necessary to postselect quantum states based on the entire measurement history. Instead, it suffices to do postselection based on a few numbers, i.e. the ``estimators", which are calculated directly from the measurement record. We showed this both analytically and by performing numerical experiments that mimic actual experimental conditions, where experimenters only have access to the measurement record and to a single copy of the conditional quantum state. By studying the structure of the estimators, we identified the presence of a memory time, indicating that only the measurement record close to the observation time is relevant for the recovery of nonlinear observables. Furthermore, we also demonstrated that the estimators can be determined based only on experimentally accessible data and/or on knowledge of the unconditional dynamics, although in our bosonic model they can be calculated analytically. In the case of the lattice model, we also showed that the estimators are governed by a coherence length, which can be tuned to be much larger than the lattice spacing, in which case a continuum description becomes appropriate. At distances larger than the lattice spacing but shorter than the coherence length the estimators display power-law behaviour.

We believe that these results open up many questions. To begin with, it is not clear whether this methodology can be extended to more complicated systems, where analytical solutions are no longer available. This could be analyzed by studying systems with increasing levels of complexity: gaussian bosons~\cite{Minoguchi2022}$\to$ few spins~\cite{lin2023}$\to$ dilute extended systems~\cite{Jin2024}, etc. The implementations described in Section~\ref{sec:ExpImp} also suggest experimentally relevant generalizations, such as the inclusion of Kerr nonlinearities in ciruit-QED (leading to a Bose-Hubbard Hamiltonian~\cite{Ma2019}) or the full nonlinear spin dynamics in cavity-QED. 

We do not expect nonlinear observables in volume law phases to be captured by any simple estimation procedure, but it may be feasible to do so in area law phases, where measurements wash out the information, presumably leading to finite correlation times and lengths. The procedure outlined at the end of section~\ref{sec:SSite:FilterExperimental} may be implementable in those cases, and an evaluation of its efficacy/efficiency is something that we defer to future work. It is also known that there are measurement-induced transitions between area law phases~\cite{Nahum2020,Ippoliti2021,Sang2021}, so it is interesting to ask whether any estimation prescription would work close to the critical point. It is also not clear what the role of the measurement record would be if the transition is second order and displays conformal symmetry, although it is possible to imagine that there might exist a mathematical description that incorporates the fluctuating nature of the record on equal footing to the fluctuations of the degrees of freedom of the system.

\section{Acknowledgements}
We thank Ana Maria Rey, Aaron Friedman, Yuxin Wang, Jayameenakshi Venkatraman and Eric Y. Song for helpful discussions and feedback on this manuscript. 
D.B. was supported by a Simons Investigator Award (Grant No.
511029). D.B. and M.P.A.F. were supported by the Simons collaboration on Ultra-Quantum Matter (UQM) which is funded by grants from the Simons Foundation (Grant No. 651440, 651457).
M.P.A.F. is also supported by a Quantum Interactive Dynamics grant from the William M. Keck Foundation. We acknowledge additional funding support from the National Science Foundation under Grant Number PFC PHY-2317149  (Physics Frontier Center). D.B. acknowledges the hospitality of the KITP while parts of this work were completed. This research was supported in part by grant NSF PHY-2309135 to the Kavli Institute for Theoretical Physics (KITP).

\bibliography{library}
\appendix

\onecolumngrid
\section{Single site gaussian dynamics}\label{app:SSite:Gaussian}
In this Appendix we begin from the stochastic Schr\"{o}dinger equation
\begin{equationS}\label{app:SSEq}
    d\hat{\rho}&=-i\left[h_0\hat{a}^\dagger\hat{a},\hat{\rho}\right]\,dt+\Gamma\left(\hat{x}\hat{\rho}\,\hat{x}-\frac{1}{2}\{\hat{x}^2,\hat{\rho}\}\right)\,dt+\sqrt{\Gamma}\Big(\hat{x}\hat{\rho}+\hat{\rho}\hat{x}-2\braket{\hat{x}}\hat{\rho}\Big)\,dW
\end{equationS}
to obtain Eq.~(\ref{eqn:SSite:Covs}) and Eq.~(\ref{eqn:SSite:Means}) under the assumption that the quantum state is gaussian. Using the cyclic property of the trace, we have that the change of any operator $\hat{O}$ after a time step, i.e. $d\braket{\hat{O}}\equiv \mathrm{Tr}(\hat{O}d\hat{\rho})$, satisfies
\begin{equationS}\label{app:SHEq}
    d\braket{\hat{O}}&=-i\left\langle\left[\hat{O},h_0\hat{a}^\dagger\hat{a}\right]+\Gamma\left(\hat{x}\hat{O}\hat{x}-\frac{1}{2}\{\hat{x}^2,\hat{O}\}\right)\right\rangle\,dt+\sqrt{\Gamma}\Big(\braket{\hat{x}\hat{O}}+\braket{\hat{O}\hat{x}}-2\braket{\hat{x}}\braket{\hat{O}}\Big)\,dW.
\end{equationS}
In particular, for $\hat{O}=\hat{x}$ we can derive
\begin{equation}
    d\braket{\hat{x}}=h_0\braket{\hat{p}}dt+2\sqrt{\Gamma}\underbrace{\left(\braket{\hat{x}^2}-\braket{\hat{x}}^2\right)}_{v_x}dW.
\end{equation}
By replacing $dW=dI-2\sqrt{\Gamma}dt$ we arrive at the first row of Eq.~(\ref{eqn:SSite:Means}). Similar manipulations lead to the equation for $d\braket{\hat{p}}$. To get equations for the covariances, we first compute the equation for $d\braket{\hat{x}^2}$ using Eq.~(\ref{app:SHEq}) with $\hat{O}=\hat{x}^2$:
\begin{equation}
    d\braket{\hat{x}^2}=h_0\braket{\hat{x}\hat{p}+\hat{p}\hat{x}}dt+2\sqrt{\Gamma}\left(\braket{\hat{x}^3}-\braket{\hat{x}}\braket{\hat{x}^2}\right)\,dW.
\end{equation}
We also need the equation for the square of the mean:
\begin{equation}
    d\left(\braket{\hat{x}}^2\right)=\Big(\braket{\hat{x}}+d\braket{\hat{x}}\Big)^2-\braket{\hat{x}}^2=2\braket{\hat{x}}d\braket{\hat{x}}+(d\braket{\hat{x}})^2,
\end{equation}
taking care of using the Ito rules $dW^2=dt$ and $dW\,dt=0$ to arrive at the right equation of motion
\begin{equationS}
    d\left(\braket{\hat{x}}^2\right)&=2h_0 \braket{\hat{x}}\braket{\hat{p}}dt+4\Gamma v_x^2 \overbrace{dt}^{dW^2}+4\sqrt{\Gamma}\left(\braket{\hat{x}}\braket{\hat{x}^2}-\braket{\hat{x}}^3\right)\,dW,
\end{equationS}
and we are explicitly indicating the term that arises from the Ito rule $dW^2=dt$. From this we construct the equation for $dv_x=d\left(\braket{\hat{x}^2}-\braket{\hat{x}}^2\right)=d\braket{\hat{x}^2}-d\left(\braket{\hat{x}}^2\right)$
\begin{equationS}
    dv_x&=2h_0 u\,dt-4\Gamma v_x^2\,dt+2\sqrt{\Gamma}\left(\braket{\hat{x}^3}-3\braket{\hat{x}^2}\braket{\hat{x}}+2\braket{\hat{x}}^3\right)\, dW,
\end{equationS}
where we have introduced $u=\braket{\{\hat{x},\hat{p}\}/2}-\braket{\hat{x}}\braket{\hat{p}}$ as in the main text. If we define the fluctuation $\Delta\hat{x}=\hat{x}-\braket{\hat{x}}$, the stochastic term in the previous equation becomes $2\sqrt{\Gamma}\braket{\Delta\hat{x}^3}\,dW$, which is $0$ for a gaussian state, leading to the first line of Eq.~(\ref{eqn:SSite:Covs}). Similar manipulations lead to the rest of Eq.~(\ref{eqn:SSite:Covs}). Notice that the nonlinear term can be traced to $dW^2=dt$ and is thus only present when measurements are included. The full set of equations for the covariances are 
\begin{equationS}
    \dot{v}_x&=2h_0 u-4\Gamma v_x^2\\
    \dot{v}_p&=-2h_0 u+\Gamma-4\Gamma u^2\\
     \dot{u}&=h_0(v_p-v_x)-4\Gamma u v_x,
\end{equationS}
and their steady state solutions are given by
\begin{equationS}\label{app:CovsSSite}
    v_x^{\infty}&=\frac{\sqrt{\frac{h_0}{8\Gamma}}}{\left(\frac{h_0}{4\Gamma}+\sqrt{\left(\frac{h_0}{4\Gamma}\right)^2+\frac{1}{4}}\right)^{1/2}}\\
    v_p^{\infty}&=\sqrt{\frac{2\Gamma}{h_0}}\times\frac{\sqrt{\left(\frac{h_0}{4\Gamma}\right)^2+\frac{1}{4}}}{\left(\frac{h_0}{4\Gamma}+\sqrt{\left(\frac{h_0}{4\Gamma}\right)^2+\frac{1}{4}}\right)^{1/2}}\\
    u^{\infty}&=\frac{1/4}{\frac{h_0}{4\Gamma}+\sqrt{\left(\frac{h_0}{4\Gamma}\right)^2+\frac{1}{4}}}
\end{equationS}

\section{Non-gaussian states}\label{app:SSite:NonGaussian}
Here we integrate the Stochastic Schrodinger Equation for pure states
\begin{equationS}
    d\ket{\psi}&=\left[-ih_0\hat{a}^\dagger\hat{a}-\frac{\Gamma}{2}(\hat{x}-\braket{\hat{x}})^2\right]\ket{\psi}\,dt+\sqrt{\Gamma}\big(\,\hat{x}-\braket{\hat{x}}\big)\ket{\psi}dW.
\end{equationS}
We rewrite this equation using the Ito rules as 
\begin{equationS}
    d\ket{\psi}&=\left[-ih_0\hat{a}^\dagger\hat{a}-\Gamma(\hat{x}-\braket{\hat{x}})^2\right]\ket{\psi}\,dt+\sqrt{\Gamma}\big(\,\hat{x}-\braket{\hat{x}}\big)\ket{\psi}dW+\frac{\Gamma}{2}\big(\,\hat{x}-\braket{\hat{x}}\big)^2\ket{\psi}dW^2
\end{equationS}
and replace $dW=dI-2\sqrt{\Gamma}\braket{\hat{x}}dt$
\begin{equationS}
    d\ket{\psi}&=\left[-ih_0\hat{a}^\dagger\hat{a}-\Gamma\hat{x}^2+\Gamma\braket{\hat{x}}^2\right]\ket{\psi}\,dt+\sqrt{\Gamma}\big(\,\hat{x}-\braket{\hat{x}}\big)\ket{\psi}dI+\frac{\Gamma}{2}\big(\,\hat{x}-\braket{\hat{x}}\big)^2\ket{\psi}dI^2
\end{equationS}
noting that $dW^2=dI^2=dt$. We extract the quadratic operators by defining $\ket{\phi}$ according to
\begin{equation}
    \ket{\psi}=\exp\left[\underbrace{\left(-ih_0\hat{a}^\dagger\hat{a}-\Gamma\hat{x}^2\right)}_{\hat{Q}}t\right]\ket{\phi},
\end{equation}
which leads to
\begin{equationS}
    d\ket{\phi}&=\Gamma\braket{\hat{x}}^2\ket{\phi}\,dt+\sqrt{\Gamma}\big(\,\hat{x}_t-\braket{\hat{x}}\big)\ket{\psi}dI+\frac{\Gamma}{2}\big(\,\hat{x}_t-\braket{\hat{x}}\big)^2\ket{\psi}dI^2,
\end{equationS}
and we have defined $\hat{x}_t= e^{-\hat{Q}t}\hat{x}e^{\hat{Q}t}$. Note that the expectation value $\braket{\hat{x}}=\braket{\psi|\hat{x}|\psi}=\braket{\phi|e^{\hat{Q}^\dagger t}e^{\hat{Q}t}\hat{x}_t|\phi}\neq \braket{\phi|\hat{x}_t|\phi}$ is time-dependent, but is just a c-number. This last equation can be integrated to give
\begin{equation}
    \ket{\phi}=e^{\Gamma\int_0^t \braket{\hat{x}}^2\,ds}\mathcal{T} \exp\left[\sqrt{\Gamma}\int_0^t (\hat{x}_s-\braket{\hat{x}})\,dI(s)\right]\ket{\psi}_0,
\end{equation}
where $\mathcal{T}$ is the time-ordering operator (note that the $dI^2$ terms are necessary to have a consistent expansion of the exponential to first order in $dt$, which includes a second order term in $dI$ because of the Ito rules). Furthermore, since $\hat{Q}$ is quadratic in boson operators then $\hat{x}_t$ is linear in $\hat{x},\hat{p}$ and we can thus express the fully evolved quantum state, up to a normalization factor (that depends on $dI$), as
\begin{equation}
    \ket{\psi}\propto e^{\hat{Q}t}\exp\left[\sqrt{\Gamma}\int_0^t\hat{x}_s\,dI(s)\right]\ket{\psi}_0,
\end{equation}
with no $\mathcal{T}$. We can calculate $\hat{x}_s$ directly
\begin{equation}
    \hat{x}_s=\frac{1}{2}\left(\hat{x}+\frac{h_0\hat{p}}{F}\right) e^{ Fs}+\frac{1}{2}\left(\hat{x}-\frac{h_0\hat{p}}{F}\right) e^{- Fs},
\end{equation}
where $F=\sqrt{-h_0^2+2i\Gamma h_0}$ is chosen to have a positive real part. Again, we can separate the two terms using Glauber's formula, which just contributes a c-number, yielding
\begin{equationS}
    \ket{\psi}\propto e^{\hat{Q}t}&\times\exp\left[\frac{\sqrt{\Gamma}}{2}\left(\hat{x}+\frac{h_0\hat{p}}{F}\right) \int_0^t e^{Fs}\,dI(s)\right]\times \exp\left[\frac{\sqrt{\Gamma}}{2}\left(\hat{x}-\frac{h_0\hat{p}}{F}\right) \int_0^t e^{-Fs}\,dI(s)\right]\ket{\psi}_0.
\end{equationS}
The leftmost exponential can be pushed past the middle one noting that $e^{\hat{Q}t}\left(\hat{x}+h_0\hat{p}/F\right)e^{-\hat{Q}t}=e^{-Ft}\left(\hat{x}+h_0\hat{p}/F\right)$ to arrive at
\begin{equationS}
    \ket{\psi}\propto &\exp\left[\frac{\sqrt{\Gamma}}{2}\left(\hat{x}+\frac{h_0\hat{p}}{F}\right) \int_0^t e^{-F(t-s)}\,dI(s)\right]\times e^{\hat{Q}t}\times \exp\left[\frac{\sqrt{\Gamma}}{2}\left(\hat{x}-\frac{h_0\hat{p}}{F}\right) \int_0^t e^{-Fs}\,dI(s)\right]\ket{\psi}_0.
\end{equationS}
This is the same as Eq.~(\ref{eqn:SSite:Non-GaussianEvolution}) once we identify
\begin{equation}\label{eqn:app:F}
    F=\sqrt{-h_0^2+2i\Gamma h_0}=\frac{1}{\tau}+i\Gamma h_0\tau,
\end{equation}
with
\begin{equation}
    \frac{1}{\tau^2}=\frac{h_0\Gamma ^2}{\frac{h_0}{2}+\sqrt{\frac{h_0^2}{4}+\Gamma^2}}
\end{equation}
This choice of operator ordering is motivated by the requirement that the arguments of the exponentials be bounded in time. In any other order there would be terms that blow up exponentially in time. 

Finally, let us show that the $e^{\hat{Q}t}$ factor projects the system onto a gaussian state of zero mean and steady state covariances given by Eq.~(\ref{app:CovsSSite}). First of all, consider the right eigenvector $\ket{\psi_R}$ of $\hat{Q}$ with eigenvalue $\lambda$ and its projection onto $\bra{\psi_R}$
\begin{equation}
    -ih_0\braket{\psi_R|\hat{a}^\dagger\hat{a}|\psi_R}-\Gamma \braket{\psi_R|\hat{x}^2|\psi_R}=\lambda \braket{\psi_R|\psi_R}.
\end{equation}
This equation implies that $\mathrm{Re}(\lambda)\leq 0$, and so at long times $e^{\hat{Q}t}$ will project any state onto the right eigenvectors whose eigenvalues have a real part as close to zero 
as possible. In particular, by expanding $\hat{a}^\dagger\hat{a}=(\hat{x}^2+\hat{p}^2-1)/2$, $\hat{Q}$ can be rewritten as
\begin{equation}
    \hat{Q}=\frac{ih_0}{2}-\frac{1}{2}\begin{pmatrix}
        \hat{x}&\hat{p}
    \end{pmatrix}\begin{pmatrix}
        ih_0+2\Gamma&0\\
        0&ih_0
    \end{pmatrix}\begin{pmatrix}
        \hat{x}\\ \hat{p}
    \end{pmatrix}
\end{equation}
We now do a similarity transformation 
\begin{equation}
    e^{-\hat{S}}\begin{pmatrix}
        \hat{x}\\
        \hat{p}
    \end{pmatrix}e^{\hat{S}}=\begin{pmatrix}
        e^{\alpha}\hat{x}\\
        e^{-\alpha}\hat{p}
    \end{pmatrix},
\end{equation}
where $\hat{S}=-i\alpha(\hat{x}\hat{p}+\hat{p}\hat{x})/2$ and $\alpha$ is a complex parameter. Choosing $e^{-2\alpha}=\frac{F}{ih_0}$ [with $F$ given by Eq.~(\ref{eqn:app:F})] leads to
\begin{equationS}
    e^{-\hat{S}}\hat{Q}e^{\hat{S}}&=\frac{ih_0}{2}-\frac{F}{2}\left(\hat{x}^2+\hat{p}^2\right)\\
    &=\frac{ih_0-F}{2}-F\hat{a}^\dagger\hat{a}.
\end{equationS}
At long times and for any state $\ket{\psi}$ we have generically that $e^{\hat{Q}t}\ket{\psi}\sim e^{\hat{S}}\ket{0}$ (where $\ket{0}$ is the vacuum state) since $e^{\hat{S}}\ket{0}$ has the smallest decay constant. We can readily obtain the wavefunction of $\ket{\psi_{\infty}}\propto e^{\hat{S}}\ket{0}$ using
\begin{equation}
    e^{\hat{S}}\hat{a}\ket{0}=\frac{1}{\sqrt{2}}(\hat{x}e^{-\alpha}+ie^{\alpha}\hat{p})e^{\hat{S}}\ket{0}=0,
\end{equation}
and projecting onto $\hat{x}$ eigenstates, yielding
\begin{equation}
    \braket{x|\psi_{\infty}}\propto \exp\left(-\frac{e^{-2\alpha}\hat{x}^2}{2}\right).
\end{equation}
By direct integration, we have that
\begin{equationS}
    \braket{\psi_\infty|\hat{x}^2|\psi_\infty}&=\frac{1}{e^{-2\alpha}+e^{-2\bar{\alpha}}}=\frac{1}{2\Gamma\tau}=v_{x}^{\infty}\\
    \braket{\psi_\infty|\hat{p}^2|\psi_\infty}&=\frac{e^{-2\alpha-2\bar{\alpha}}}{e^{-2\alpha}+e^{-2\bar{\alpha}}}=v_p^{\infty}\\
    \braket{\psi_\infty|\{\hat{x},\hat{p}\}|\psi_\infty}&=\frac{i(e^{-2\alpha}-e^{-2\bar{\alpha}})}{e^{-2\alpha}+e^{-2\bar{\alpha}}}=\frac{1}{\Gamma h_0\tau^2}=2u^{\infty},
\end{equationS}
as defined in Eq.~(\ref{app:CovsSSite}), which is what we set out to prove.

\section{Record-record correlation function}\label{app:SSite:RecordRecord}
Here we calculate 
\begin{equation}\label{app:record-recordEquation}
     R(s-s')ds ds'=\lim_{T\to\infty}\frac{1}{T}\int_{-T/2}^{T/2}dt\,\overline{dI(t-s)dI(t-s')}.
\end{equation}
In general~\cite{Wiseman1993,zoller1997quantumnoisequantumoptics},
\begin{align}\begin{split}
    \overline{\frac{dI(t_1)}{dt_1}\frac{dI(t_2)}{dt_2}}&=\delta(t_1-t_2)+4\Gamma\, \mathrm{Re}\left\{\mathrm{Tr}\left[\hat{x}e^{\mathcal{L}|t_1-t_2|}(\hat{x}e^{\mathcal{L}(t_{\text{min}}-t_0)}\hat{\rho}_0)\right]\right\},
\end{split}\end{align}
where $t_{\text{min}}$ is the smallest between $t_1>t_0$ and $t_2>t_0$, and $\hat{\rho}_0$ is the state of the system at the time $t_0$. For the purposes of Eq.~(\ref{app:record-recordEquation}) we take $t_0=-T/2$. In the case of Eq.~(\ref{eqn:SSite:SSE}), we can solve for this correlator analytically, yielding
\begin{align}\begin{split}\label{app:record-recordEquation2}
    \overline{\frac{dI(t_1)}{dt_1}\frac{dI(t_2)}{dt_2}}&=\delta(\tau)+4\Gamma\left\{\frac{\Gamma}{4h_0}\sin(h_0|\tau|)-\frac{\Gamma|\tau|}{4}\cos(h_0\tau)\right\}\\
    &+4\Gamma \left\{\frac{\braket{\hat{x}^2-\hat{p}^2}_{-T/2}}{2}\cos\left[h_0(t_1+t_2+T)\right]\right\}+4\Gamma \left\{\left(\frac{\braket{\{\hat{x},\hat{p}\}}_{-T/2}}{2}-\frac{\Gamma}{4h_0}\right)\sin\left[h_0(t_1+t_2+T)\right]\right\}\\
    &+4\Gamma\left\{\frac{\braket{\hat{x}^2+\hat{p}^2}_{-T/2}}{2}+\frac{\Gamma(t_1+t_2+T)}{4}\right\}\cos(h_0\tau),
\end{split}\end{align}
where $\tau=t_2-t_1$. Plugging Eq.~(\ref{app:record-recordEquation2}) into Eq.~(\ref{app:record-recordEquation}), we find that the first line is unaffected, the second line yields 0, while the third line needs some more analysis:
\begin{equationS}\label{eqn:app:Rfinalcasi}
    R&=\delta(s-s')+4\Gamma\left\{\frac{\Gamma}{4h_0}\sin(h_0|s-s'|)-\frac{\Gamma|s-s'|}{4}\cos(h_0|s-s'|)\right\}\\ &+4\Gamma\left\{\frac{\braket{\hat{x}^2+\hat{p}^2}_{-T/2}}{2}+\frac{\Gamma(T-s-s')}{4}\right\}\cos(h_0|s-s'|).
\end{equationS}
The last line of the previous equation is not time-translationally invariant and diverges as $T\to\infty$ but we can omit it by recalling that $R$ is meant to be integrated against the filter $f$ in Eq.~(\ref{eqn:SSite:FilterSolution}) and hoping that the diverging terms will cancel against a similar contribution on the right-hand side of Eq.~(\ref{eqn:SSite:FilterSolution}). In fact, we have
\begin{equationS}
    &\overline{\braket{\hat{x}(t)}dI(t-s)}=2\sqrt{\Gamma}\,\mathrm{Re}\left\{\mathrm{Tr}\left[\hat{x}e^{\mathcal{L}s}(\hat{x}e^{\mathcal{L}(t-s-t_0)}\rho_{0})\right]\right\}\,dt.
\end{equationS}
We have already calculated this correlator. Setting $t_0=-T/2$ and integrating over $t$, we get, at large $T$
\begin{equationS}
   \frac{1}{T} \int_{-T/2}^{T/2}&\overline{\braket{\hat{x}(t)}dI(t-s)}=2\sqrt{\Gamma}\left[\frac{\Gamma}{4h_0}\sin(h_0s)\right]+2\sqrt{\Gamma}\left[\frac{\braket{\hat{x}^2+\hat{p}^2}_{-T/2}}{2}+\frac{\Gamma(T-2s)}{4}\right]\cos(h_0 s).
\end{equationS}
To be able to cancel the $\propto T$ terms in Eq.~(\ref{eqn:SSite:FilterSolution}) we thus need 
\begin{equation}
    \int_0^{\infty} f(s')\cos[h_0(s-s')]\,ds'=\frac{1}{2\sqrt{\Gamma}}\cos(h_0 s),
\end{equation}
which is equivalent to two conditions on $f(s)$:
\begin{equationS}\label{eqn:App:FilterConditions}
    \int_0^{\infty} f(s)\cos(h_0s)\,ds&=\frac{1}{2\sqrt{\Gamma}}\\
    \int_0^{\infty} f(s)\sin(h_0s)\,ds&=0.
\end{equationS}
These conditions can be used to massage Eq.~(\ref{eqn:SSite:FilterSolution}) into a more amicable form
\begin{equation}\label{eqn:App:FilterSolution}
    f(s)+\int_{0}^{\infty} r_*(s-s')f(s')\,ds'=0
\end{equation}
with $r_*$ given by
\begin{align}\begin{split}
    r_*(s)&=-2\theta(-s)\left[\frac{\Gamma^2}{h_0}\sin(h_0s)-\Gamma^2 s\cos(h_0s)\right],
\end{split}\end{align}
and where $\theta(s)$ is a Heaviside theta function. Differentiating with respect to $s$ various times leads to a differential equation for $f(s)$:
\begin{equation}
    \left[\left(\frac{d^2}{ds^2}+h_0^2\right)^2+4\Gamma^2h_0^2\right]f(s)=0
\end{equation}
This equation together with Eq.~(\ref{eqn:App:FilterConditions}) fix $f(s)$ uniquely to be the function given in Eq.~(\ref{eqn:SSite:FilterEquation}). 

\section{General filter}\label{app:GeneralFilter}
In this Appendix we solve the minimization problem posed by Eq.~(\ref{eqn:SSite:CostFunctionalGeneral}) for a generic system. To frame the problem, we copy here the associated cost functional:
\begin{equation}
    C=\overline{\big[\braket{\hat{O}(t)}-O_{\text{est}}(dI)\big]^2},
\end{equation}
but we omit the time integral since we assume that the system has evolved for a long enough time that the unconditional observables are time-independent, we are choosing to estimate the generic operator $\hat{O}$ and we also assume that we are measuring a generic operator $\hat{M}$ (not necessarily $\hat{x}$). If we expand the quadratic term we have
\begin{equation}
    C=\overline{\braket{\hat{O}(t)}^2}-2\overline{\braket{\hat{O}(t)}O_{\text{est}}(dI)}+\overline{O_{\text{est}}(dI)O_{\text{est}}(dI)}.
\end{equation}
We now need to minimize over $x_{\text{est}}(dI)$. To parameterize this functional, we express it as a functional Fourier transform:
\begin{equation}
    O_{\text{est}}(dI)=\int F(\zeta) e^{i\int \zeta(s) dI(s)}\,\mathcal{D}\zeta 
\end{equation}
and then perform minimization over $F(\zeta)$. For clarity, we remark that this involves functional-\textit{al} differentiation: we are differentiating with respect to the functional $F$, not with respect to $\zeta$, which are integration variables. Hence each $F(\zeta_{t_0},\zeta_{t_1},...)$ for each different value of $\zeta$ at each different time is an independent variable. Then
\begin{equationS}
    \frac{\delta}{\delta F(\zeta)} \left[\overline{O_{\text{est}}(dI)x_{\text{est}}(dI)}\right]&=2\,\overline{O_{\text{est}}(dI) e^{i\int \zeta(s)dI(s)}}\\
     \frac{\delta}{\delta F(\zeta)} \left[\overline{\braket{\hat{O}(t)}O_{\text{est}}(dI)}\right]&=\overline{\braket{\hat{O}(t)}e^{i\int \zeta(s)dI(s)}},
\end{equationS}
so that
\begin{equation}
    \overline{O_{\text{est}}(dI) e^{i\int \zeta(s)dI(s)}}=\overline{\braket{\hat{O}(t)}e^{i\int \zeta(s)dI(s)}}.
\end{equation}
We multiply by $e^{-i\int \zeta(s)dI'(s)}$ and functionally integrate over $\zeta(s)$ 
\begin{equation}
    \int D\zeta e^{-i\int \zeta(s) dI'(s)}\left(\,\overline{O_{\text{est}}(dI) e^{i\int \zeta(s)dI(s)}}\right)=\int D\zeta e^{-i\int \zeta(s) dI'(s)}\left(\,\overline{\braket{\hat{O}(t)}e^{i\int \zeta(s)dI(s)}}\right).
\end{equation}
Since the integral over $\zeta$ leads to a delta functional that picks the measurement realization $dI(s)=dI'(s)$ we can pull out the $O_{\text{est}}(dI)$ from the overline (average over measurement realizations) if we replace $dI$ by $dI'$. Hence
\begin{equation}
    O_{\text{est}}(dI')=\left[\int D\zeta e^{-i\int \zeta(s) dI'(s)}\left(\,\overline{e^{i\int \zeta(s)dI(s)}}\right)\right]^{-1}\times \int D\zeta e^{-i\int \zeta(s) dI'(s)}\left(\,\overline{\braket{\hat{O}(t)}e^{i\int \zeta(s)dI(s)}}\right).
\end{equation}
The averages of $e^{-i\int \zeta(s) dI(s)}$ are related to the characteristic density operator~\cite{zoller1997quantumnoisequantumoptics} at time $t$
\begin{equation}
    \hat{\chi}(\zeta)=\overline{\hat{\rho}_t\,e^{i\int \zeta(s) dI(s)}},
\end{equation}
in terms of which we can write the estimator as
\begin{equation}\label{eqn:app:GeneralFilter}
\boxed{O_{\text{est}}(dI')=\left\{\int D\zeta e^{-i\int \zeta(s) dI'(s)}\mathrm{Tr}\left[\hat{\chi}(\zeta)\right]\right\}^{-1}\times \int D\zeta e^{-i\int \zeta(s) dI'(s)}\mathrm{Tr}\left[\hat{O}\hat{\chi}(\zeta)\right]}.
\end{equation}
The characteristic density operator can be written as~\cite{zoller1997quantumnoisequantumoptics}
\begin{equation}
    \hat{\chi}(\zeta)=\exp\left[-\frac{1}{2}\int_0^t\zeta(s)^2\,ds\right]\mathcal{T}\left\{\exp\left[\int_0^t\left(\mathcal{L}_0+\mathcal{L}_{\hat{M}}+i\zeta(s)\mathcal{M}\right)\,ds\right]\right\}\hat{\rho}_0,
\end{equation}
where $\mathcal{T}$ is the time-ordering superoperator, $\hat{\rho}_0$ is the initial state of the system,
\begin{equationS}
    \mathcal{M}\hat{\rho}&=\sqrt{\Gamma}\left(\hat{M}\hat{\rho}+\hat{\rho}\hat{M}^\dagger\right)\\
    \mathcal{L}_{\hat{M}}&=\Gamma\left(\hat{M}\hat{\rho}\hat{M}^\dagger-\frac{1}{2}\{\hat{M}^\dagger\hat{M},\hat{\rho}\}\right)
\end{equationS}
and $\mathcal{L}_0$ describes any other dynamical process ocurring simultaneously with the continuous measurement. Note that $\hat{M}$ is not necessarily hermitian, as may happen in the case of homodyne detection. Although this is a general solution to the estimation problem, the presence of the functional integrals make it unusable in practice unless some more information is provided. In Eq.~(\ref{eqn:app:GeneralFilter}) only the functional Fourier transform of $\hat{\chi}(\zeta)$ appears:
\begin{equation}\label{eqn:app:rhodi0}
    \hat{\rho}(dI)=\int D\zeta e^{-i\int \zeta(s) dI(s)}\mathrm{Tr}\left[\hat{\chi}(\zeta)\right],
\end{equation}
which can be done exactly because $\zeta(s)$ only appears quadratically. Then
\begin{equation}\label{eqn:app:rhodI}
    \hat{\rho}(dI)=\exp\left[-\frac{1}{2}\int_0^t\left(\frac{dI(s)}{ds}\right)^2\,ds\right]\mathcal{T}\left\{\exp\left[\int_0^t \left(\mathcal{L}_0\,ds+ dI(s)\mathcal{M}+\mathcal{N}_{\hat{M}}ds\right)\right]\right\}\hat{\rho}_0,
\end{equation}
where
\begin{equation}
    \mathcal{N}_{\hat{M}}\hat{\rho}=-\frac{\Gamma}{2}(\hat{M}+\hat{M}^{\dagger})\hat{M}\hat{\rho}-\frac{\Gamma}{2}\hat{\rho}\hat{M}^\dagger(\hat{M}+\hat{M}^\dagger).
\end{equation}
Then the estimator is given by
\begin{equation}
    \hat{O}_{\text{est}}(dI)=\frac{\mathrm{Tr}[\hat{O}\hat{\rho}(dI)]}{\mathrm{Tr}[\hat{\rho}(dI)]},
\end{equation}
and $\hat{\rho}(dI)$ is thus the conditional density operator. This can be verified by replacing $dI(s)=\sqrt{\Gamma}\braket{\hat{M}+\hat{M}^\dagger}\,ds+dW(s)$, differentiating with respect to time (taking into account the Ito rule), renormalizing the state and checking that we recover the stochastic Schr\"odinger equation. In the presence of only Hamiltonian evolution, this can be reframed for pure states as
\begin{equation}
    \ket{\psi(dI)}\propto \mathcal{T}\exp\left\{\int_0^t\left(-i\hat{H}\,ds+dI(s)\sqrt{\Gamma}\hat{M}-\frac{\Gamma}{2} \left(\hat{M}+\hat{M}^\dagger\right)\hat{M}\,ds\right)\right\}\ket{\psi_0}.
\end{equation}
Note also that if we integrate either Eq.~(\ref{eqn:app:rhodi0}) or Eq.~(\ref{eqn:app:rhodI}) with respect to $dI(s)$, corresponding to an average over measurement realizations, we obtain
\begin{equation}
    \underline{\hat{\rho}}=\int\hat{\rho}(dI)\,\mathcal{D}\dot{I}=\hat{\chi}(0)=\exp\left[\left(\mathcal{L}_0+\mathcal{L}_{\hat{M}}\right)t\right]\hat{\rho}_0,
\end{equation}
i.e. the unconditional Lindblad evolution. From this perspective, the measurement record ($\dot{I}=dI/ds$) can be interpreted as a Hubbard-Stratonovich decoupling field for the $\hat{M}\hat{\rho}\hat{M}^\dagger$ term in the Lindblad equation. The different possible decoupling choices then correspond to the different possible measurements that give rise to the same unconditional evolution.

\end{document}